%% file: ms.tex
\PassOptionsToPackage{hyperfootnotes=false}{hyperref}
\documentclass[twocolumn]{aastex701}
\usepackage{amsmath}

\input{numbers}

\newcommand{\Ha}{H$\alpha$}
\newcommand{\Hb}{H$\beta$}
\newcommand{\kms}{\ensuremath{{\rm km\,s^{-1}}}}
\newcommand{\ergs}{\ensuremath{{\rm erg\,s^{-1}}}}
\newcommand{\sigint}{\sigma_{\rm int}}
\newcommand{\LHa}{L_{\rm H\alpha}}
\newcommand{\Lbol}{L_{\rm bol}}
\newcommand{\Lph}{L_{\rm ph}}
\newcommand{\Rph}{R_{\rm ph}}
\newcommand{\Teff}{T_{\rm eff}}
\newcommand{\vK}{v_{\rm K}}
\newcommand{\Mvir}{M_{\rm vir}}
\providecommand{\Msun}{M_\odot}
\newcommand{\BX}{BX26}

\shorttitle{Broad-line widths of little red dots}
\shortauthors{Wang}
\hypersetup{pdftitle={A Tight Broad-line Width-Luminosity Relation in Little Red Dots and Its Consequence for Virial Black Hole Masses},pdfauthor={Haoyu Wang}}

\begin{document}

\title{A Tight Broad-line Width--Luminosity Relation in Little Red Dots and Its Consequence for Virial Black Hole Masses}

\author[orcid=0009-0005-9117-0842,gname=Haoyu,sname=Wang]{Haoyu Wang}
\affiliation{Independent Researcher}
\email[show]{pkuwanghaoyu@pku.edu.cn}
\correspondingauthor{Haoyu Wang}

\begin{abstract}
JWST has uncovered compact high-redshift sources with red, $\sim 5000$~K optical continua and broad Balmer lines, known as ``little red dots'' (LRDs).
Whether these lines trace gas orbiting a massive black hole or electron scattering in a dense, Eddington-limited envelope around a lighter black hole leaves inferred masses spanning $10^{4}$--$10^{8}\,\Msun$.
We test both pictures by uniformly refitting public JWST/NIRSpec spectra of \NLRD{} LRDs, \NOther{} other JWST broad-line sources, and luminosity-matched SDSS quasars.
At fixed \Ha{} luminosity, LRD broad \Ha{} widths have an intrinsic scatter of \Fsig{}~dex ($<\FsigUp{}$~dex at 95\%), less than half that of quasars (\SsgTwo{}~dex), clustering near $\WConstMed{}$~\kms{} outside the \NBright{} brightest sources.
Because single-epoch virial masses scale as width squared, they scatter intrinsically by $<\sigMvir{}$~dex at fixed \Ha{} luminosity, making virial masses and mass functions nearly deterministic restatements of luminosity regardless of the underlying physics.
Compared with the Keplerian speed $\vK$ at the photosphere of an Eddington-limited envelope, from each source's measured continuum luminosity and temperature, observed widths are \ratLo{}--\ratHi{}\,$\vK$ ($2\text{--}3\times$ standard expectations for virial factors $f_{\rm vir}\approx0.5\text{--}1$), with a quarter of the flux above escape speed and no detected continuum-luminosity dependence at fixed \Ha{} luminosity.
Bound orbits outside a spherical photosphere thus require sub-Eddington optical luminosities ($\leq\lamVirLo{}\text{--}\lamIsoHi{}\,L_{\rm Edd}$, a requirement specific to spherical photospheres), whereas an Eddington or super-Eddington spherical envelope requires uniform $2\text{--}3\times$ non-gravitational broadening within the \Ha{}-emitting gas.
\end{abstract}


\section{Introduction} \label{sec:intro}

Deep JWST imaging and spectroscopy have uncovered an abundant population of compact, red sources with broad Balmer emission lines (${\rm FWHM}\gtrsim1000$~\kms{}) at $z\sim3$--8, known as ``little red dots'' \citep[LRDs;][]{Matthee2024,Greene2024,Kocevski2025}.
Their comoving number densities ($10^{-5}$--$10^{-4}$~Mpc$^{-3}$) exceed UV-selected quasar luminosity functions at the same epochs by one to two orders of magnitude before declining steeply below $z\simeq4$ \citep{Kocevski2025,Akins2025,Inayoshi2025}.
Unlike classical type~1 quasars, LRDs combine unresolved optical morphologies with ``V-shaped'' rest-frame UV-to-optical continua inflecting near the Balmer limit, frequently accompanied by Balmer breaks too strong for evolved stars at realistic densities \citep{Setton2025Vshape,InayoshiMaiolino2025,deGraaffRuby2025,NaiduBHStar2026,Hviding2025} and narrow, often blueshifted Balmer absorption lines \citep{Matthee2024,InayoshiMaiolino2025,DEugenio2026,Shangguan2026,ChenCH2026}.
They also exhibit flat rest-frame near- to mid-infrared spectral energy distributions and far-infrared non-detections that disfavor hot-torus or cold-starburst dust re-emission \citep{Williams2024,Akins2025,Setton2025IR,Greene2026,Iani2026}, alongside extreme X-ray weakness in deep \emph{Chandra} stacks \citep{Ananna2024,Yue2024} and weak or undetected UV/optical variability \citep{Kokubo2025,Zhang2025,Furtak2025,Burke2025}.

Although non-AGN models invoke host kinematics in ultra-compact stellar cores \citep{Baggen2024}, cluster tidal disruption events \citep{Bellovary2025}, or primordial supermassive stars \citep{NandalLoeb2026}, gas-phase Balmer breaks \citep{deGraaffRuby2025,NaiduBHStar2026}, narrow forbidden lines, and lensed dynamical kinematics \citep{Juodzbalis2026} favor accreting black holes in most broad-line LRDs.
Even among AGN models, however, weak dust re-emission and red optical continua resembling $\Teff\simeq5000$~K thermal radiation \citep{InayoshiHo,deGraaff2026} leave inferred black hole masses spanning $10^{4}$--$10^{8}\,\Msun$.
Standard single-epoch virial estimators give $M_{\rm BH}\sim10^{7}$--$10^{8}\,\Msun$ \citep{Matthee2024,Taylor2025,Sun2026}, implying high black hole-to-stellar mass ratios, straining So{\l}tan radiative-efficiency budgets \citep{Soltan1982} under large bolometric corrections \citep{InayoshiIchikawa2024,Akins2025,Greene2026}, and motivating rapid growth via heavy seeds, super-Eddington episodes, or dark-matter-assisted collapse \citep{Inayoshi2025,Gu2026}.
Three families of AGN continuum models span this mass range: (i) anisotropic or equatorially reddened disks that obscure the UV/X-ray core while exposing a virial broad-line region \citep[BLR;][]{Madau2026a,Madau2026c}; (ii) flattened cool ($\sim2000$--$5000$~K) thermal disks with mild dust attenuation ($A_V\lesssim1$~mag), whether from a gravitationally unstable ($Q\sim1$) outer disk paired with an inner standard disk \citep{ZhangCX2026}, the $T\simeq5000$~K inner rim of a circumbinary disk around a compact ($\lesssim10^{3}$ Schwarzschild radii) massive black hole binary \citep{InayoshiChen2026}, or the outer ``Hayashi limit'' of a self-gravitating disk \citep{ChenYX2026}; and (iii) dense, Eddington- or super-Eddington spherical envelopes (``black hole stars'') forming $\sim5000$~K photospheres around $10^{6}$--$10^{7}\,\Msun$ black holes \citep{NaiduBHStar2026,LiuHP2025,BegelmanDexter2025,Kido2025,deGraaff2026}, or $10^{4}$--$10^{5}\,\Msun$ when combined with Balmer-absorption speeds or supercritical outflows \citep{Sun2026,Naidu2026,Curtis2026,LiuFengHo2026}.

At the center of this $10^{4}$--$10^{8}\,\Msun$ spread is what the broad Balmer line width physically traces.
In a gravitational reading, widths trace bound gas in the central potential: spatially resolved kinematics in the lensed LRD A2744-QSO1 match its single-epoch virial mass \citep{Juodzbalis2026}, differing \Ha{}, \Hb{}, and Pa$\beta$ profiles in a luminous LRD disfavor simple Thomson scattering \citep{Brazzini2025}, and radial stratification in a dense virial BLR can produce Lorentzian or exponential wings and centrally peaked Balmer decrements without scattering \citep{Madau2026b,Scholtz2026,ChenCH2026}.
In a scattering reading, an intrinsic Doppler or magnetospheric core of width $\lesssim\vK$ near the photosphere \citep{Lin2025,Inayoshi2026,Burke2025,TakasaoInayoshi2026} traverses dense ionized gas where Thomson, resonance, or Raman scattering forms broad wings set by electron temperature and optical depth \citep{Rusakov2026,Chang2025,Sneppen2026,Matthee2026}; related supercritical-outflow models match 15 of 18 LRD widths within a factor of 1.5 from photospheric luminosity and temperature \citep{LiuFengHo2026}.

Thus far, these mechanisms have been tested mainly through profile fits to individual objects---where core-plus-wing and stratified virial profiles can be degenerate---or composite stacks that average out source-to-source scatter.
Here we pose a two-part population test: how much do LRD broad \Ha{} widths scatter intrinsically at fixed luminosity, and how do unsplit widths compare and scale with each source's measured Eddington-photosphere Keplerian speed?
In lower-redshift quasars, small width dispersion at fixed luminosity constrains BLR geometry, inclination selection, and Eddington-ratio spread \citep{Kollmeier2006,Fine2008,Shen2008} while showing that single-epoch virial masses largely track luminosity \citep{Croom2011,Assef2012}.
Although \citet{BaccusXu2026} compiled JWST broad-line widths against luminosity, the intrinsic scatter of the LRD relation---isolated from profile-shape systematics, unmodeled instrumental variance, and selection effects---has not been measured or confronted with individual photospheric velocity scales.

We uniformly refit public JWST/NIRSpec grating spectra of \NLRD{} LRDs, \NOther{} other JWST broad-line sources, and luminosity-matched SDSS quasars across six width definitions (Section~\ref{sec:data}), and measure the slope and intrinsic scatter alongside the catalog of \citet{BaccusXu2026} (Section~\ref{sec:results}).
Section~\ref{sec:tests} calibrates unmodeled error floors from duplicate exposures and tests profile, narrow-core, and selection biases via injection--recovery simulations.
Section~\ref{sec:physics} combines the scatter ceiling with $\vK(\Rph)$ to constrain spherical orbital and scattering-envelope models; Section~\ref{sec:discussion} examines flattened thermal disks, binaries, non-AGN scenarios, virial-mass demographics, and observational tests; and Section~\ref{sec:summary} summarizes our conclusions.
We adopt the Planck 2018 flat $\Lambda$CDM cosmology \citep[$H_0=\Hnot{}$~\kms~Mpc$^{-1}$, $\Omega_{\rm m}=\OmM{}$;][]{Planck2020}; catalogs using $H_0=\Hcat{}$~\kms~Mpc$^{-1}$ and $\Omega_{\rm m}=\OmCat{}$ differ by at most \dCosmoL{}~dex in $\log L$ (\dCosmoVK{}~dex in $\log\vK$) across our redshifts.
Logarithms are base 10; unless marked as one-sided 95\% upper limits, intervals are equal-tailed 95\% credible intervals.

\section{Data and Measurements} \label{sec:data}

\subsection{Spectra and Line Fits} \label{sec:fit}

We use public medium- and high-resolution NIRSpec grating spectra from the DAWN JWST Archive \citep[DJA;][]{DJAHeintz2023,DJAdeGraaff2024}, reduced with \texttt{msaexp} \citep{msaexp} including geometric point-source slit-loss calibration, and add a \errFloor{}\% pixel-error floor in quadrature to the pipeline uncertainties.
Merging published LRD and broad-line catalogs \citep[including][]{Taylor2025,Kocevski2026,Hviding2025,deGraaff2026} within \matchRadius{} yields \NSources{} objects with \NSpectra{} grating spectra, \NSpecHa{} covering \Ha{}.

Within $|\Delta v|\leq\fitWin{}$~\kms{} around \Ha{} (blueward of \ion{He}{1}~$\lambda6678$ and [\ion{S}{2}]~$\lambda\lambda6716,6731$), we fit a linear continuum, narrow \Ha{} and [\ion{N}{2}]~$\lambda\lambda6548,6583$ fixed to 1:3 in flux with one Gaussian FWHM (\wnarLo{}--\wnarHi{}~\kms{}), and a broad \Ha{} component (\wbroadLo{}--\wbroadHi{}~\kms{}).
Models are convolved with the nominal wavelength-dependent grating line-spread function \citep[LSF;][]{Jakobsen2022} and integrated over pixels, so fitted widths are intrinsic.
A multiplicative Gaussian optical-depth screen $\exp[-\tau(v)]$ on \Ha{} ($\tau_0\leq\tauMax{}$, FWHM \wabsLo{}--\wabsHi{}~\kms{}) is included when it lowers the Bayesian information criterion (BIC) by more than \cutBIC{} (\NAbsLRD{} LRDs and \NAbsOther{} other sources in the final sample).
Line luminosities $\LHa$ refer to the unabsorbed broad component without dust correction.

\subsection{Six Measures of Width} \label{sec:widths}

We fit the broad component in turn as a single Gaussian, a Lorentzian, an exponential, and a sum of two Gaussians, and also recover a non-negative piecewise-linear profile regularized by cross-validation.
These fits define six widths: the FWHM of the four parametric models (of the summed profile for two Gaussians) plus the nonparametric FWHM and interquartile width $W_{50}=v_{75}-v_{25}$ within $\pm\fitWin{}$~\kms{}.
Each width is paired with the luminosity and covariance of its own fit, from the full parameter Jacobian for parametric models and from \nBootNp{} pixel-noise bootstrap realizations at fixed narrow-line, continuum, and absorption subtraction for the nonparametric profile.

Most LRD profiles have stronger wings than a Gaussian: the Lorentzian or exponential has the lowest BIC in \NBICwing{} of the \NLRD{} LRDs, beating the Gaussian by $\Delta{\rm BIC}>\cutBIC{}$ in \NBICsix{} (including all \NBIChiSN{} with broad-line ${\rm S/N}>20$, versus \NBICsixOth{} of the \NOther{} other sources).
Assigning each LRD its own BIC-preferred parametric form (\NBICgauss{} Gaussian, \NBIClor{} Lorentzian, \NBICexp{} exponential) mixes width definitions whose zero points differ by a median of \dWge{}~dex between Gaussian and exponential fits, inflating the apparent scatter to \DsigBic{}~dex.
We therefore apply one uniform measure at a time across all sources.
We adopt the single-Gaussian FWHM as our baseline summary of line width because it is well-defined for every object, does not compress parent scatter on winged profiles at the observed widths (Section~\ref{sec:injtests}), and gives a scatter similar to every other uniform single-component or nonparametric measure (Section~\ref{sec:definitions}).

\subsection{The Sample} \label{sec:sample}

A fit passes quality control if the broad component has ${\rm S/N}\geq\cutSN{}$, beats the narrow-only model by $\Delta{\rm BIC}\geq\cutBIC{}$, exceeds the instrumental FWHM without hitting a broad-width or absorption bound, has reduced $\chi^2<\cutChi{}$, and converged.
We froze these rules on the single-Gaussian fits before computing population scatters, after inspecting initial fits to set the $\chi^2<\cutChi{}$ threshold and to restrict bound rejection to broad-width and absorption parameters.
Relaxing both adjustments leaves $\sigint=\sigIncA{}$~dex, and the rule history, measurements, covariances, and reproduction scripts are archived at \url{https://github.com/NianSi-AI/lrd-width-luminosity}.
For objects with multiple passing spectra we retain the spectrum with the highest broad-line S/N as the primary spectrum.
Classifications as LRD or non-LRD come from \citet{Taylor2025}, \citet{Kocevski2026}, and \citet{Hviding2025}; sources with missing or conflicting labels are excluded.

The resulting sample comprises \NLRD{} LRDs and \NOther{} other broad-line sources in the EGS and UDS fields, the two fields covered by those three catalogs (Table~\ref{tab:sample}).
The LRDs span $z=\zLRDmin{}$--$\zLRDmax{}$, $\log(\LHa/\ergs)=\LLRDmin{}$--$\LLRDmax{}$ (median \LLRDmed{}), and ${\rm FWHM}=\WLRDmin{}$--\WLRDmax{}~\kms{} (median \WLRDmed{}), with median broad-line S/N of \SNLRDmed{}; the \NOther{} other sources span $z=\zOthmin{}$--$\zOthmax{}$ with median S/N of \SNOthmed{}.
None is gravitationally lensed, and sample membership is held fixed across all six width measures.
CEERS-746 and Skyfire-299 lie \sepPair{}\arcsec{} apart at the same redshift ($z=5.623$) in separate slitlets of one MSA mask; we treat them as distinct objects.

\input{tables/Table1_stub}

\subsection{Comparison Data} \label{sec:bx}

\emph{Published JWST fits.}
\citet[][hereafter \BX{}]{BaccusXu2026} report single-Gaussian broad-line fits after sideband continuum subtraction for \NX{} NIRSpec records (nominal cut ${\rm FWHM}>\XWcut{}$~\kms{}, retaining previously identified broad-line AGN below it).
Their \NXLRD{} LRD records correspond to \NXLRDuniq{} unique objects on the sky, and \NXOvRec{} records overlap our sample; their measurements are independent fits to the same public DJA/\texttt{msaexp} reductions.
Because \BX{} convert broad \Ha{} flux (dereddened from total \Ha/\Hb{} where \Hb{} is covered, or by their sample-median reddening otherwise) into $\Lbol$ using a constant bolometric correction within each class, their width scatter at fixed $\Lbol$ is directly comparable to ours whereas their slope is not.
As their comparison group we take the \NXCtrl{} non-LRD records at $z>\XzCut{}$ in the LRD $\Lbol$ range.

\emph{Low-redshift quasars.}
From the SDSS DR7 catalog of \citet{Shen2011} we select all \NSDSS{} quasars ($z=\zSDSSmin{}$--$\zSDSSmax{}$) with broad \Ha{} ${\rm FWHM}>\HvidingCut{}$~\kms{}, line-luminosity ${\rm S/N}>\cutSN{}$, and $\LHa$ in our LRD range.
Because catalog widths come from multi-Gaussian decompositions after host and \ion{Fe}{2} subtraction, we retrieved SDSS DR17 spectra for a random \NSsgDraw{} of these quasars and refitted them around \Ha{} (where optical \ion{Fe}{2} is weak) with the identical procedure and quality rules of Sections~\ref{sec:fit} and \ref{sec:sample}; \NSsg{} pass.

\subsection{Statistical Model} \label{sec:model}

For each sample we fit
\begin{equation} \label{eq:model}
\log {\rm FWHM} = a + b\,\log\!\left(\frac{L}{L_0}\right) + \varepsilon, \qquad \varepsilon\sim\mathcal{N}(0,\sigint^2),
\end{equation}
with $L_0=10^{43}$~\ergs{} for $\LHa$ and $10^{45}$~\ergs{} for $\Lbol$.
For source $i$ with logarithmic luminosity and width $(x_i,y_i)$ and measurement covariance $C_i$, the residual variance is
\begin{equation} \label{eq:var}
V_i=\sigint^2 + C_{yy,i} + b^2 C_{xx,i} - 2b\,C_{xy,i} .
\end{equation}
We adopt uniform priors in $a$, $b\in(-1,1)$, and $\sigint\in(0,0.6)$~dex, marginalize $a$ analytically on a fine grid, and compute class differences by convolving the marginal posteriors of $\sigint$.
Replacing this conditional likelihood with a hierarchical Gaussian prior on the true luminosities gives identical LRD parameters ($b=\hierB{}$, $\sigint=\hierSig{}$~dex).

\section{Results} \label{sec:results}

\subsection{The Width--Luminosity Relation} \label{sec:relation}

\begin{figure*}
\includegraphics[width=\textwidth]{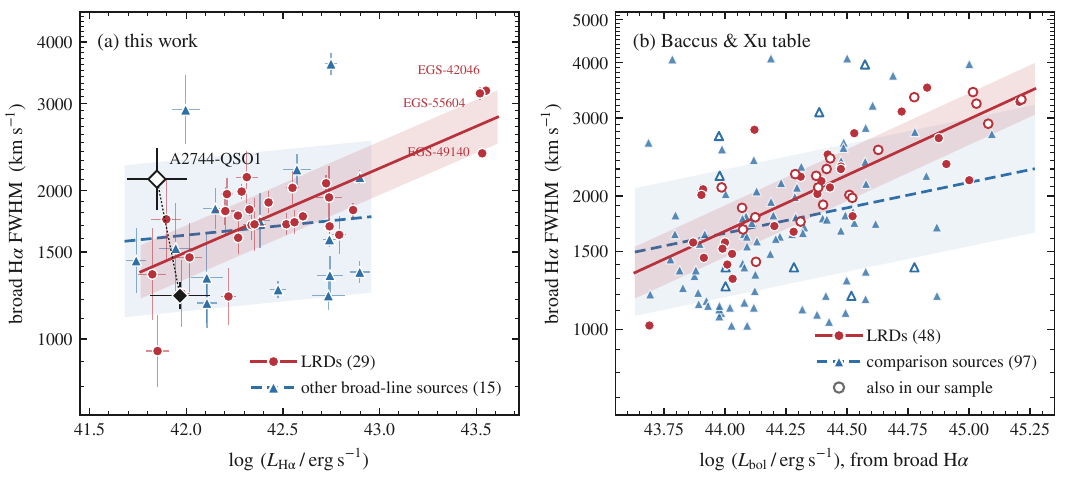}
\caption{Broad \Ha{} width against luminosity.
(a) Our sample: \NLRD{} LRDs (circles) and \NOther{} other broad-line sources (triangles), fitted with one single-Gaussian procedure ($1\sigma$ error bars); labels mark the \NBright{} most luminous LRDs, which set the slope.
Diamonds show A2744-QSO1, which is not in our sample, corrected for lensing magnification, with the narrow line free (open) or fixed at \wnarDef{}~\kms{} (filled; Section~\ref{sec:variants}).
(b) The table of \citet{BaccusXu2026}, using $\Lbol$ scaled from broad \Ha{} and comparison sources at $z>\XzCut{}$; open symbols mark records shared with our sample, with horizontal jitter up to 0.03~dex to separate rounded tabulated luminosities, and error bars are omitted for clarity.
Lines and shaded bands show posterior-median relations and $\pm\sigint$ under formal errors.
\label{fig:relation}}
\end{figure*}

Under formal fitting covariances, the \NLRD{} LRDs follow
\begin{equation} \label{eq:relation}
\log\frac{\rm FWHM}{\kms} = \Fa{} + \Fb{}\,\log\frac{\LHa}{10^{43}\,\ergs}
\end{equation}
with slope $b=\Fb{}$ (\Fblo{}--\Fbhi{}) and intrinsic scatter $\sigint=\Fsig{}$~dex (\Fsiglo{}--\Fsighi{}, one-sided 95\% upper limit \FsigUp{}~dex; Figure~\ref{fig:relation}a and Table~\ref{tab:fits}).
Even with measurement errors ignored ($C=0$), the raw residual standard deviation is \Frms{}~dex.
In the \BX{} table the \NXLRD{} LRD records have $\sigint=\Xsig{}$~dex (\Xsiglo{}--\Xsighi{}) and slope \Xb{} (\Xblo{}--\Xbhi{}) against $\Lbol$ (Figure~\ref{fig:relation}b), or $\sigint=\XsigDedup{}$~dex (\XsigDeduplo{}--\XsigDeduphi{}) after deduplicating to the \NXLRDuniq{} unique objects.

The positive slope in our sample rests on its \NBright{} brightest LRDs near $\LHa\approx10^{\LBright{}}$~\ergs{}, separated from the remaining \NNoTop{} below $10^{\LRestMax{}}$~\ergs{}; removing those \NBright{} objects lowers the slope to \FbNoTop{} (\FbNoToplo{}--\FbNoTophi{}) while leaving $\sigint=\FsigNoTop{}$~dex (\FsigNoToplo{}--\FsigNoTophi{}).
The remaining \NNoTop{} LRDs, spanning $\log\LHa=\LConstMin{}$--\LConstMax{}, are consistent with a constant width: fixing $b=0$ gives $\sigint=\FsigConst{}$~dex (\FsigConstlo{}--\FsigConsthi{}) around a median FWHM of \WConstMed{}~\kms{}.
The slope also depends on how the line core is split between narrow and broad components ($b=\FbNarFix{}$--\Fb{}; Section~\ref{sec:variants}) and on the luminosity definition: the \BX{} LRDs span $\Lbol=10^{\XLmin{}}$--$10^{\XLmax{}}$~\ergs{} without a gap and give \Xb{} (\Xblo{}--\Xbhi{}) against dereddened $\Lbol$.
None of our conclusions relies on its value.
A2744-QSO1 (diamonds in Figure~\ref{fig:relation}a) is not in our sample and is discussed in Section~\ref{sec:variants}.

\subsection{Six Definitions of Width} \label{sec:definitions}

\begin{figure*}
\centering\includegraphics[width=\textwidth]{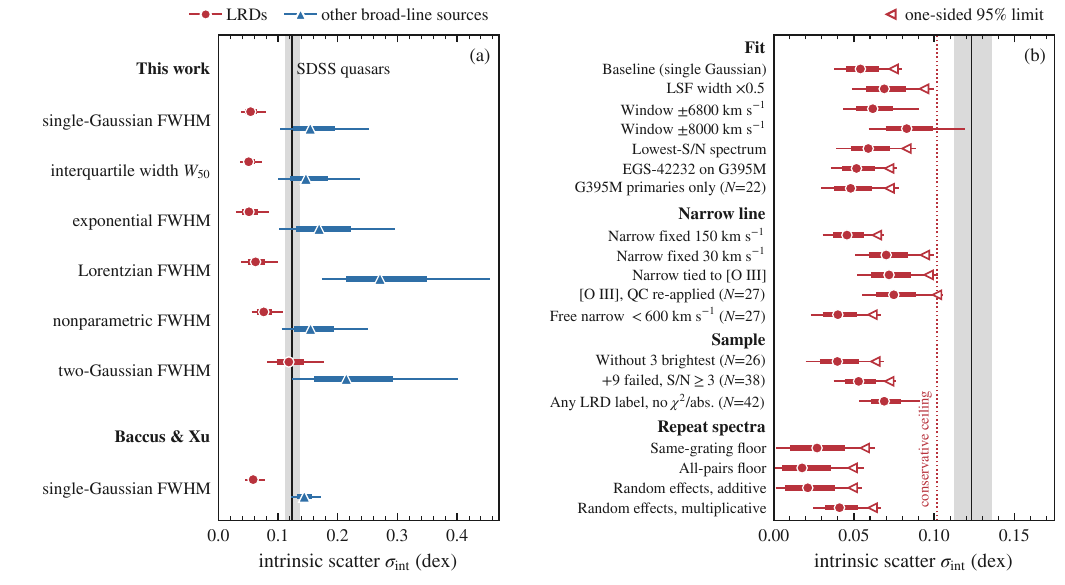}
\caption{Intrinsic scatter of broad \Ha{} widths at fixed luminosity under formal fitting errors, for (a) six width definitions in our sample and single-Gaussian fits in \citet{BaccusXu2026}, and (b) analysis variants of the LRD single-Gaussian FWHM (Section~\ref{sec:tests}).
Points mark posterior medians; thick and thin bars show 68\% and 95\% intervals, and open triangles mark one-sided 95\% upper limits where full posteriors are evaluated.
Repeat-spectrum rows include unmodeled variance calibrated from duplicate spectra.
The dotted line marks the conservative ceiling of \FsigCons{}~dex, and in both panels the vertical line with gray band shows the \NSsg{} luminosity-matched SDSS quasars refitted with our single-Gaussian procedure.
\label{fig:forest}}
\end{figure*}

Five of the six uniform width measures give posterior-median $\sigint=\sigFiveLo{}$--$\sigFiveHi{}$~dex for the LRDs under formal errors, and in every one the 95\% interval of $\Delta\sigint$ relative to the other sources excludes zero (Figure~\ref{fig:forest} and Table~\ref{tab:fits}).
Only the two-Gaussian sum shows larger scatter (\DsigTg{}~dex, \DsigTglo{}--\DsigTghi{}), because a separate core and wing are identifiable in just \NCoreWing{} of the \NFrozen{} spectra.

\subsection{Other Broad-line Sources and Quasars} \label{sec:others}

Other JWST broad-line sources scatter by \Fsigo{}~dex (\Fsigolo{}--\Fsigohi{}) in our sample and \Xsigo{}~dex (\Xsigolo{}--\Xsigohi{}) in \BX{}, giving $\Delta\sigint\equiv\sigma_{\rm int,LRD}-\sigma_{\rm int,other}=\Fdsig{}$ and \Xdsig{}~dex (Table~\ref{tab:fits}).
Our own control group has only \NOther{} objects, and dropping its two widest members (\WOutA{} and \WOutB{}~\kms{}) shifts $\Delta\sigint$ to \FdsigDroplo{} to \FdsigDrophi{}~dex; the statistical weight of the comparison therefore comes from the \BX{} sample and from SDSS quasars.
Inflating every formal error by factors of \inflLo{}--\inflHi{}, or adding \inflFlrLo{}--\inflFlrHi{}~dex in quadrature, keeps $\Delta\sigint$ at \dsigInflLo{} to \dsigInflHi{}~dex with 95\% intervals excluding zero in both samples, so the contrast does not arise from underestimated errors.
The comparison class is heterogeneous, however, and may include outflow-dominated or low-S/N broad components.

Luminosity-matched SDSS DR7 quasars scatter by \Ssig{}~dex in the multi-Gaussian catalog widths of \citet{Shen2011} (\SsgCat{}~dex on the refitted subset) and by \Ssg{}~dex (\Ssglo{}--\Ssghi{}, slope \Ssgb{}) when refitted with our single-Gaussian pipeline (\NSsg{} quasars).
At the median LRD luminosity ($\LHa=10^{\LLRDmed{}}$~\ergs{}), the refitted quasar relation is not only \sdssRatSame{} times more scattered in log width than the LRD locus under formal errors, its median width is also \sdssWoff{}~dex wider (\WsdssMedL{} versus \WlrdMedL{}~\kms{}).
Removing the $\chi^2$ rule adds \NSsgChi{} quasars and gives \SsgNoChi{}~dex.
The $\pm\fitWin{}$~\kms{} window spans only about $\pm1.5$ FWHM for these quasars (those rejected for low S/N have median FWHM \WSdssSNfail{}~\kms{}), so truncated wings, if anything, lower the SDSS scatter and width offset.

\input{tables/Table2_fits}

\section{Error Calibration and Selection Tests} \label{sec:tests}

\subsection{Duplicate-Spectrum Error Calibration} \label{sec:repeats}

Formal Jacobian covariances capture pixel noise conditional on the parametric profile, omitting slit-loss, grating-calibration, and line-shape systematics.
Nine LRDs have multiple spectra (\NSpecPassLRD{} across all \NLRD{} LRDs, forming \NPairsAll{} intra-object pairs; \NPairsBigObj{} from RUBIES-EGS-42232) that independently pass Section~\ref{sec:sample}.
Normalized by formal width uncertainties, pairwise $\log{\rm FWHM}$ differences have rms \rmsDupSameG{} across the \NPairsSameG{} same-grating pairs and \rmsDupAll{} across all \NPairsAll{} pairs (\rmsDupBest{} across \NPairsBest{} primary--secondary pairs).
Adding the corresponding $+\flrDupSameG{}$ and $+\flrDupAll{}$~dex per-spectrum floors in quadrature lowers $\sigint$ to \FsigFlrSame{} and \FsigFlrAll{}~dex ($<\FsigFlrSameUp{}$ and $<\FsigFlrAllUp{}$~dex; $<\flrSliceUpLo{}$--$\flrSliceUpHi{}$~dex under equal-object or leave-one-out reweightings) while preserving $\Delta\sigint<0$ (Table~\ref{tab:fits}).
Joint random-effects fits to all \NSpecPassLRD{} spectra at fixed primary luminosity likewise give $\sigint<\reAllUp{}$~dex with an additive floor (median \reAddFloor{}~dex; $<\reDropBigUp{}$~dex keeping one spectrum of RUBIES-EGS-42232), or $<\reMultUp{}$~dex when unmodeled variance scales with $C_{yy}$ (median factor \reMultScale{}).
Repeats also show that G395H widths are narrower than G395M in all \NGHobj{} overlapping LRDs (median \dWGH{}~dex; $\offGH{}\pm\offGHerr{}$~dex as a free random-effects offset, leaving the $\sigint$ limit unchanged; the \NGM{} G395M-primary LRDs alone give $\sigint=\FsigGM{}$~dex, $<\FsigGMUp{}$~dex) and that retaining the highest-S/N spectrum favors narrower fits by \dWPrim{}~dex on average.
Because the \NSNfif{} LRDs with ${\rm S/N}\geq15$ already have a raw residual standard deviation of \rmsSNfif{}~dex---below the $+\flrDupAll{}$~dex all-pairs floor---repeat spectra bound $\sigint$ to \reAllUp{}--\reMultUp{}~dex depending on the error model (consistent with zero only for an additive floor), and we use formal errors below.

\subsection{Injection--Recovery and Pre-cut Completeness} \label{sec:injtests}

In simulations, neither profile mismatch nor the quality cuts manufacture this tightness.
In \NReal{}-realization injection--recovery suites built on the noise and LSF templates of the \NLRD{} surviving LRD spectra (Appendix~\ref{app:injections}, Table~\ref{tab:injections}), single-Gaussian fits to Gaussian, exponential, and Lorentzian lines with $\sigma_{\rm parent}=\simA{}$~dex yield $\sigint\leq\Fsig{}$~dex in \hitTenAll{} of \nullDenThree{} trials combined.
Luminosity-independent ($b=0$) winged injections produce a slope as steep as \Fb{} in \hitSlopeNull{} of \nullDen{} trials.
The quality cuts are also insensitive to width: on all \NParentInBand{} in-band pre-cut templates (below), injections shifted $+\passShift{}$~dex above the locus pass \passRatLo{}--\passRatHi{} times as often as those on it, and $\sigma_{\rm parent}=\simA{}$~dex yields $\sigint\leq\Fsig{}$~dex in \hitParTenG{} of \NReal{} Gaussian and \hitParTenE{} of \NReal{} exponential trials after the full quality cuts.
Including the rejected spectra changes little.
Of the \NParentClean{} consistently labeled LRDs with a grating \Ha{} fit before quality cuts, \NParentInBand{} have \Ha{} in band (longward of the order-blocking filter cut-on) and \NParentFail{} of these fail Section~\ref{sec:sample}.
Adding the \NFailSNthree{} with broad-line ${\rm S/N}\geq\cutSNthree{}$ gives $\sigint=\FsigSNthree{}$~dex ($<\FsigSNthreeUp{}$~dex at 95\%; $N=\NLRDSNthree{}$, Table~\ref{tab:fits}), although \NFailSNthreeWeak{} of the nine have $\Delta{\rm BIC}<\cutBIC{}$ against the narrow-only model; \NFailPosUDS{} of them lie $+\resFailPosLo{}$ to $+\resFailPosHi{}$~dex above Equation~(\ref{eq:relation}) at ${\rm S/N}=\failPosSNlo{}$--$\failPosSNhi{}$, with or without a fixed narrow line.

\subsection{Narrow-Core Decomposition, A2744-QSO1, and Analysis Variants} \label{sec:variants}

The single-Gaussian width depends on how the line core is split between the narrow and broad components.
With its FWHM free, the narrow Gaussian sits at the \wnarLo{}~\kms{} lower bound in \NNarThirty{} of the \NLRD{} LRDs, and in \NNarBoundLRD{} of the \NBright{} luminous LRDs (none of the \NOther{} other sources) it reaches the \wnarHi{}~\kms{} upper bound and absorbs the core, widening the broad Gaussian from $\simeq\WNarBoundAfter{}$ to $\simeq\WNarBoundBefore{}$~\kms{}; [\ion{O}{3}]~$\lambda5007$ is unresolved (at the \wnarLo{}~\kms{} bound) in \NOIIIUnres{} of the \NOIIIMeas{} LRDs where it is measured.
Anchoring the narrow \Ha{} FWHM to [\ion{O}{3}], the standard choice, changes individual widths by \dWNarRms{}~dex rms (up to \dWNarMax{}~dex) and gives $\sigint=\FsigNarOIII{}$~dex ($<\FsigNarOIIIUp{}$~dex), or \FsigOIIIqc{}~dex ($<\FsigOIIIqcUp{}$~dex; $N=\NOIIIqc{}$) after re-applying the absorption-bound rule of Section~\ref{sec:sample}.
Fixing every narrow FWHM at \wnarLo{} or \wnarDef{}~\kms{} gives \FsigNarMin{} and \FsigNarFix{}~dex, and excluding the \NNarBoundLRD{} bound objects gives \FsigNarCut{}~dex.
These variants lower the slope to \FbNarFix{}--\FbNarOIII{}, and $\Delta\sigint$ excludes zero in every case (Table~\ref{tab:fits}); they hold membership fixed, although whether a broad component is required also depends on the split.

A2744-QSO1, which has a dynamical black hole mass \citep{Juodzbalis2026}, is not in our sample: it lies in the Abell~2744 field, which none of our three classification catalogs covers, and with a free narrow line its broad component has ${\rm S/N}=\QSOsn{}$, below our cut.
In that fit the narrow Gaussian reaches the \wnarHi{}~\kms{} bound and leaves a \QSOW{}~\kms{} broad Gaussian $\QSOresCor{}\pm\QSOresErr{}$~dex above Equation~(\ref{eq:relation}) for the magnification $\mu=\QSOmu{}\pm\QSOmuErr{}$ of image~B \citep{Furtak2025}.
Its broad profile has \QSOcompNar{} and \QSOcompBrd{}~\kms{} components \citep{DEugenio2026,Juodzbalis2026}; with the narrow-line FWHM fixed at \wnarDef{}~\kms{}, its single-Gaussian FWHM is \QSOWfix{}~\kms{} (${\rm S/N}=\QSOsnFix{}$, $\QSOresFix{}\pm\QSOresFixErr{}$~dex from Equation~(\ref{eq:relation})), but this fit is not preferred over a single \wnarHi{}~\kms{} narrow component without a broad line ($\Delta{\rm BIC}=\QSOdBICfree{}$), so its single-Gaussian width is poorly defined.
Adding it to the [\ion{O}{3}]-anchored fit leaves $\sigint$ unchanged at \FsigNarOIIIqso{}~dex ($N=\NLRDqso{}$).

Remaining sample and fitting checks behave similarly (Figure~\ref{fig:forest}, Table~\ref{tab:fits}).
Our faintest LRDs sit \floorDex{}~dex above the \HvidingCut{}~\kms{} catalog cut.
Restricting to ${\rm S/N}\geq20$ (\NSNtwenty{} LRDs, \sigSNtwenty{}~dex), halving the LSF (\FsigLsf{}~dex, $<\FsigLsfUp{}$~dex), choosing each object's lowest-S/N passing spectrum (\FsigLowSN{}~dex, $<\FsigLowSNUp{}$~dex), or replacing the 3.91~$\mu$m G235M primary spectrum of RUBIES-EGS-42232 with G395M (\FsigSwap{}~dex, $<\FsigSwapUp{}$~dex) all keep 95\% limits below \FsigCons{}~dex.
Admitting any LRD label across all four input catalogs (Section~\ref{sec:fit}) gives \sigIncB{}~dex ($N=\NIncB{}$, or \sigIncD{}~dex at $N=\NIncD{}$ without the $\chi^2$ and absorption-bound cuts, driven by one $\chi^2=\chiHiB{}$ fit), and widening the fit window to $\pm\winWide{}$ or $\pm\winWider{}$~\kms{} ([\ion{S}{2}] masked) gives \FsigWin{} and \FsigWinWide{}~dex (\FsigWinWidelo{}--\FsigWinWidehi{}).
In \BX{}, the \NXNoOv{} non-overlapping LRDs give \XsigNoOv{}~dex while \NXPairs{} matched LRD--control pairs differ in residual scatter by \XdMatch{}~dex.

Across these fit, narrow-line, and sample variants (Figure~\ref{fig:forest}), one-sided 95\% limits reach \FsigCons{}~dex, set by the [\ion{O}{3}]-anchored fit with the absorption rule re-applied, and only the widest fit window extends the 95\% interval further, to \FsigWinWidehi{}~dex.
We therefore quote $\sigint<\FsigUp{}$~dex as the baseline limit and \FsigCons{}~dex as a conservative ceiling that covers these analysis choices.

\section{Physical Constraints from Widths and Photospheric Scaling} \label{sec:physics}

We now derive the quantitative constraints that the small width dispersion, the photospheric velocity scale, and the absence of detected photospheric scaling place on bound orbital motion and electron-scattering envelopes.
Throughout Sections~\ref{sec:physics} and \ref{sec:discussion}, upper bounds (``at most'' or ``$\leq$'') are evaluated at the baseline 95\% limit $\sigint=\FsigUp{}$~dex; bounds linear in $\sigint$ loosen by \narCeilLoosenPct{}\% at the conservative \FsigCons{}~dex ceiling of Section~\ref{sec:tests} and tighten by a quarter at the same-grating repeat-spectrum limit (\FsigFlrSameUp{}~dex).

\subsection{The Scatter Budget of an Orbital Reading} \label{sec:budget}

\begin{figure}
\includegraphics[width=\columnwidth]{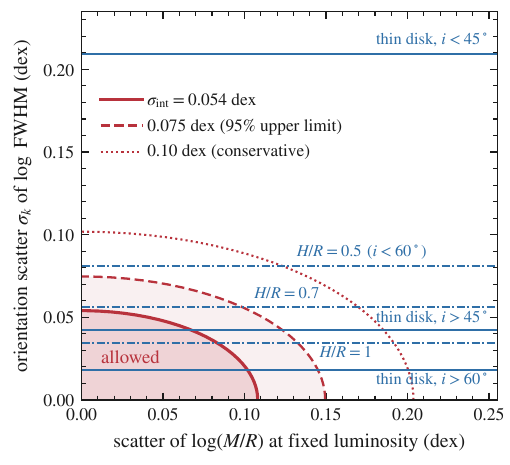}
\caption{Allowed scatter of $\log(M/R)$ and orientation factor $\log k$ ($\sigma_k=\tfrac{1}{2}\sigma_{\log f_{\rm vir}}$) at fixed $\LHa$ in an orbital reading (Equation~\ref{eq:budget}), bounded by the formal-error posterior median ($\sigint=\Fsig{}$~dex, solid), the one-sided 95\% upper limit (\FsigUp{}~dex, dashed; the same-grating repeat-spectrum 95\% limit of \FsigFlrSameUp{}~dex lies close to the solid curve), and the conservative ceiling (\FsigCons{}~dex, dotted).
Horizontal lines mark the orientation scatter of a thin disk ($i<45^\circ$, $i>45^\circ$, and $i>60^\circ$) and of velocity fields with isotropic thickness $H/R$ ($i<60^\circ$).
\label{fig:budget}}
\end{figure}

If the width traces bound orbital motion, ${\rm FWHM}=f_{\rm vir}^{-1/2}(GM/R)^{1/2}\equiv k\,(GM/R)^{1/2}$, where $M=f_{\rm vir}R\,{\rm FWHM}^{2}/G$ is the mass within emitting radius $R$ and $k=f_{\rm vir}^{-1/2}$ encodes geometry and inclination, then at fixed $\LHa$
\begin{equation} \label{eq:budget}
\sigint^{2} = \sigma_{k}^{2} + \tfrac{1}{4}\,\sigma_{M/R}^{2} = \tfrac{1}{4}\,\sigma_{\log f_{\rm vir}}^{2} + \tfrac{1}{4}\,\sigma_{M/R}^{2}
\end{equation}
when $f_{\rm vir}$ (geometry and viewing angle) is uncorrelated with $M/R$ (Figure~\ref{fig:budget}).
Hence $\log(M/R)$ at fixed $\LHa$ scatters by at most \sigLamMax{}~dex.
In local reverberation-mapped AGN, the \Hb{} BLR radius scatters by \RLlo{}--\RLhi{}~dex at fixed 5100~\AA{} luminosity, including measurement error \citep{Bentz2013}; if a scatter this large held at fixed $\LHa$ with $M$ independent of $R$, it would contribute up to \RLtermLo{}--\RLtermHi{}~dex to $\sigint$, comparable to our 95\% limit before any scatter in black hole mass or inclination enters.

Inclination narrows the budget further.
A thin disk viewed at $i<45^\circ$ has $\sigma_k=\oriThin{}$~dex, \oriThinRatio{} times the entire \FsigUp{}~dex limit, and survives at low inclination only if $k$ and $M/R$ cancel with correlation coefficient $-\rhoThin{}$.
Adding an isotropic component, ${\rm FWHM}^{2}\propto\sin^{2}i+(H/R)^{2}$ \citep{Decarli2011}, lowers $\sigma_k$ over $i<60^\circ$ to \oriHalf{}, \oriSeven{}, and \oriOne{}~dex for $H/R=0.5$, 0.7, and 1 (leaving $\sigma_{M/R}\leq\sigLamIso{}$~dex at $H/R=1$).
Alternatively, because $\sin i$ flattens near the equatorial plane, viewing LRDs through dusty equatorial gas at $45^\circ<i<90^\circ$ or $60^\circ<i<90^\circ$ \citep{Madau2026a,Madau2026c} yields $\sigma_k=\oriEdgeWide{}$ or \oriEdge{}~dex, leaving $\sigma_{M/R}\leq\sigLamEdgeWide{}$ or \sigLamEdge{}~dex.
Finally, because $\LHa$ is not corrected for dust, attenuation of broad \Ha{} by $A_{\rm H\alpha}$ magnitudes, with dispersion $\sigma_A$ at fixed intrinsic $\LHa$, shifts $\log\LHa$ by $-0.4\,A_{\rm H\alpha}$ and adds $(0.4\,b\,\sigma_A)^2$ to $\sigint^2$ unless wider lines suffer less extinction, requiring $\sigma_A\leq\sigDustMax{}$~mag at $b=\Fb{}$ (\sigDustNoTop{}~mag from the $N=\NNoTop{}$ subsample limit at $b=\FbNoTop{}$ without the \NBright{} brightest LRDs).

\subsection{The Photospheric Velocity Scale and Null Scaling Test} \label{sec:photosphere}

A spherical photosphere of luminosity $\Lph$ and temperature $\Teff$ has radius $\Rph=(\Lph/4\pi\sigma_{\rm SB}\Teff^{4})^{1/2}$.
Writing $\Lph=\lambda\,L_{\rm Edd}$ relative to the conventional ionized-hydrogen electron-scattering Eddington luminosity $L_{\rm Edd}=4\pi GM c/\kappa_{\rm es}$ ($\kappa_{\rm es}=\kappaES{}$~cm$^{2}$~g$^{-1}$; solar-composition $\kappa=\kappaSolar{}$~cm$^{2}$~g$^{-1}$ lowers speeds by \vKdrop{}\%), where $\lambda\equiv\Lph/L_{\rm Edd}\leq\Lbol/L_{\rm Edd}$ measures the photospheric component alone, the Keplerian speed at radius $R$ is $\vK\,(\lambda R/\Rph)^{-1/2}$.
Here
\begin{equation} \label{eq:vk}
\vK = \vKnorm{}~\kms\left(\frac{\Lph}{10^{45}\,\ergs}\right)^{1/4}\frac{\Teff}{5000~{\rm K}}
\end{equation}
is fixed entirely by the continuum with nothing fitted to the lines, giving
\begin{equation} \label{eq:ratio}
{\rm FWHM} = f_{\rm vir}^{-1/2}\vK\left(\frac{\lambda R}{\Rph}\right)^{-1/2} \equiv k\,\vK\left(\frac{\lambda R}{\Rph}\right)^{-1/2},
\end{equation}
with $f_{\rm vir}=k^{-2}$ as defined in Section~\ref{sec:budget}.

\begin{figure*}
\includegraphics[width=\textwidth]{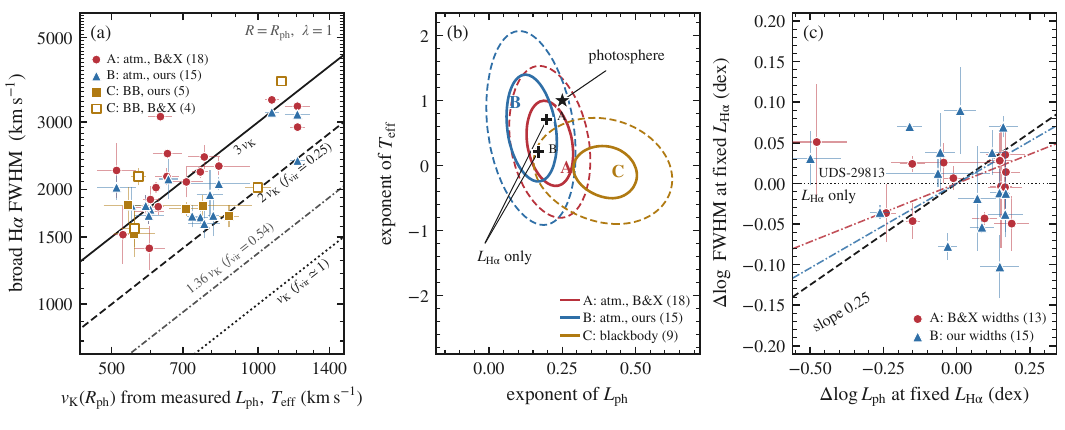}
\caption{Comparison of broad \Ha{} widths with the Eddington-photosphere Keplerian speed $\vK(\Rph)$ evaluated from each LRD's measured $\Lph$ and $\Teff$ ($\TminP{}$--\TmaxP{}~K; Equation~\ref{eq:vk}), with horizontal error bars propagating both uncertainties.
(a) Single-Gaussian FWHM against $\vK(\Rph)$ at $R=\Rph$ and $\lambda=1$ for atmosphere fits \citep[Sample~A, circles: \BX{} widths, labeled B\&X; Sample~B, triangles: our widths;][]{Curtis2026} and modified-blackbody fits \citep[Sample~C, filled squares: our widths; open squares: \BX{} widths;][]{deGraaff2026} (25 unique LRDs across the three separately analyzed samples, including 10 shared between A and B), compared with orbital virial benchmarks ${\rm FWHM}=f_{\rm vir}^{-1/2}\vK$: the 1:1 line ($f_{\rm vir}=1$, within 4\% of the empirical AGN calibration $f_{\rm vir}=1.08$; dotted), isotropic 3D Gaussian dispersion ($f_{\rm vir}=0.54$, $1.36\,\vK$, dash-dotted), extreme edge-on circular ring ($f_{\rm vir}=0.25$, $2\,\vK$, dashed), and $3\,\vK$ (solid).
(b) Joint 68\% (solid) and 95\% (dashed) posterior contours of $(b,c)$ in ${\rm FWHM}\propto\Lph^{b}\Teff^{c}$ for Samples~A, B, and C; the star marks the photospheric scaling $(0.25,1)$, while plus signs show the locus induced in A and B if widths depend on $\LHa$ alone.
(c) Partial residuals of $\log{\rm FWHM}$ against $\log\Lph$ at fixed $\LHa$ in the atmosphere samples (\NJA{} and \NJB{} sources; Appendix~\ref{app:scaling}), compared with the photospheric slope $0.25$ (dashed), the median partial slopes recovered from photospheric-scaling simulations at the measured errors (\simMedJA{} and \simMedJB{}, dash-dotted), and a pure $\LHa$ dependence (dotted); the labeled point is RUBIES-UDS-29813.
\label{fig:photo}}
\end{figure*}

Evaluating $\vK(\Rph)$ individually from each source's spectroscopically measured $(\Lph,\Teff)$, we match line widths to published atmosphere fits from \citet[][\NPA{} LRDs with \BX{} widths, Sample~A; \NPB{} with our widths, Sample~B]{Curtis2026} and modified-blackbody fits from \citet[][\NPE{} LRDs with our widths, Sample~E; \NPC{} adding \NPCbx{} \BX{} widths, Sample~C; Appendix~\ref{app:scaling}]{deGraaff2026}, spanning $\Lph=10^{\LphMin{}}$--$10^{\LphMax{}}$~\ergs{} and $\Teff=\TminP{}$--\TmaxP{}~K.
Across these samples the median single-Gaussian ${\rm FWHM}/\vK$ is \ratLo{}--\ratHi{} (Figure~\ref{fig:photo}a; $2.4$ on Sample~B when narrow \Ha{} is anchored to [\ion{O}{3}] or fixed to 150~\kms{}), with an intrinsic ratio scatter of \sqLo{}--\sqHi{}~dex in the atmosphere samples.
Whereas envelope models assign $\lesssim\vK(\Rph)$ to an intrinsic gravitational or magnetospheric core \citep{Inayoshi2026,TakasaoInayoshi2026} and invoke electron scattering (or an electron-scattered outflow; \citealt{LiuFengHo2026}) to reach the observed widths, Figure~\ref{fig:photo}a compares the unsplit broad-line FWHM directly with $\vK(\Rph)$ from the continuum alone.
Other width definitions shift the median ratio to \ratNpB{}--\ratNpE{} (nonparametric FWHM) and \ratExLo{}--\ratWingHi{} (exponential or Lorentzian FWHM; matching $f_{\rm vir}$ for the exponential profile preserves the $\sim2.5\times$ offset, Section~\ref{sec:readings}) while keeping the ratio scatter at \sqMeasLo{}--\sqMeasHi{}~dex.
Independent of parametric form or $f_{\rm vir}$, integrating each source's nonparametric profile within $\pm\fitWin{}$~\kms{} places a median \fKepNpLo{}--\fKepNpHi{}\% of the broad-\Ha{} flux above $\vK$ and \fEscNpLo{}--\fEscNpHi{}\% above the photospheric escape speed $\sqrt{2}\,\vK$ (\fEscInfLo{}--\fEscInfHi{}\% when the best-BIC profiles are extrapolated to $\pm\infty$).

Despite setting a characteristic velocity scale, the photospheric parameters show no detected influence on source-to-source width variations.
In free fits ${\rm FWHM}\propto\Lph^{b}\Teff^{c}$, $(0.25,1)$ lies on the \hpdPA{}\%--\hpdPC{}\% contours, whereas the posteriors sit directly on the locus induced by $\LHa$ alone through $\Lph$--$\LHa$ correlation (Figure~\ref{fig:photo}b; Appendix~\ref{app:scaling}).
$\log(\Lph/\LHa)$ scatters by \sdLphHaLo{}--\sdLphHaHi{}~dex across the atmosphere samples, yet at fixed $\LHa$ the partial slope on $\log\Lph$ is \bpJA{} (\bpJAlo{} to \bpJAhi{}) and \bpJB{} (\bpJBlo{} to \bpJBhi{}), flat residuals reproduced in only \simPctJA{}\% and \simPctJB{}\% of photospheric-scaling simulations (Figure~\ref{fig:photo}c).
This null test rests largely on RUBIES-UDS-29813, which lies \dLphOut{}~dex below the $\Lph$--$\LHa$ trend; without it the partial slopes become \bpJAdrop{} (\bpJAdroplo{} to \bpJAdrophi{}) and \bpJBdrop{} (\bpJBdroplo{} to \bpJBdrophi{}), consistent with both zero and the simulated photospheric medians (\simMedJA{} and \simMedJB{}).
Line luminosity alone predicts the widths of these objects to \slJA{}--\slJB{}~dex, \ratTightLo{}--\ratTightHi{} times more tightly than $\vK(\Rph)$ (\soJA{}--\soJB{}~dex).

\subsection{Spherical Orbital and Scattering Readings} \label{sec:readings}

Table~\ref{tab:readings} synthesizes what bound orbital motion outside a spherical photosphere and an Eddington-limited scattering envelope must each satisfy to match the velocity scale, small scatter, and null photospheric scaling; non-spherical thermal disks, compact binaries, and non-AGN alternatives are examined in Section~\ref{sec:alternatives}.

\input{tables/Table3_readings}

\emph{Orbital motion outside a spherical photosphere.}
Gas bound to the mass inside a spherical photosphere moves no faster than the escape speed, $\sqrt{2}\,\vK\,\lambda^{-1/2}$ at $\Rph$ and less beyond it, so the quarter of the flux beyond $\sqrt{2}\,\vK$ requires $\lambda<1$ if broad \Ha{} comes from bound orbits outside $\Rph$.
By Equation~(\ref{eq:ratio}), $\lambda R/\Rph=(k\,\vK/{\rm FWHM})^{2}={\rm FWHM}^{-2}\vK^{2}/f_{\rm vir}$, where $f_{\rm vir}=k^{-2}$ must match the profile whose width is used.
For Gaussian widths, an isotropic velocity dispersion ($f_{\rm vir}=0.54$, $k=\kIso{}$) gives $\lambda R/\Rph=\lamIsoLo{}$--\lamIsoHi{}, the standard AGN virial factor of \citet[$f_{\rm vir}=1.08$, $k=\kVir{}$]{ReinesVolonteri2015} gives \lamVirLo{}--\lamVirHi{}, and an isotropic exponential profile ($f_{\rm vir,{\rm exp}}=3.12$, $k=\kIsoExp{}$) applied to exponential widths gives \lamIsoExpLo{}--\lamIsoExpHi{}.
Even the extreme $f_{\rm vir}=0.25$ ($k=\kRing{}$) of an edge-on circular ring gives $\lambda R/\Rph=\lamRingLo{}$--\lamRingHi{} (within systematic uncertainties of unity).
For $R\geq\Rph$, typical isotropic or virial geometries thus limit the photosphere to $\lambda\leq\lamVirLo{}$--\lamIsoHi{} and place the enclosed mass at $(\mEddStdLo{}$--$\mEddHi{})\,R/\Rph$ times $M_{\rm Edd}(\Lph)$ ($\mEddLo{}\times$ for $f_{\rm vir}=0.25$, $k=\kRing{}$).
Here $\lambda$ refers to the total mass enclosed within $R$, black hole plus any interior envelope.
Spherical mass outside $R$ leaves circular speeds unchanged but deepens the potential at $\Rph$ by a fraction $\delta$ of the $M_{\rm Edd}(\Lph)$ potential ($v_{\rm esc}^{2}=2\vK^{2}[\lambda^{-1}+\delta]$), so bound gas beyond $\sqrt{2}\,\vK$ requires $\lambda<1/(1-\delta)$.
Continuum-thin gas ($\tau_{\rm ext}\leq1$, $\rho\propto r^{-2}$ to $\rOutRph{}\,\Rph$) gives $\delta=\deltaKthreeLo{}$--\deltaKthreeHi{} across our sample temperatures at $\kappa=10^{-3}$~cm$^{2}$~g$^{-1}$ and at most \deltaKfourMax{} ($\lambda<\lamEscKfourMax{}$) for $\kappa\approx10^{-4}$~cm$^{2}$~g$^{-1}$, the surface opacity outside the H$^{-}$ band in the atmosphere models of \citet[][Appendix~A]{Curtis2026}.
Achieving $\delta\sim1$ at $r\gg\Rph$ requires an exterior gas mass $M_{\rm ext}\sim(r/\Rph)\,M_{\rm Edd}(\Lph)\gg M_{\rm Edd}(\Lph)$, and continuum optical depth does not bound stellar or non-spherical mass.
An outward radiative acceleration lowers both orbital and escape speeds by $(1-\Gamma)^{1/2}$ and tightens the $\lambda$ ceiling.
If broad \Ha{} originates inside a polar cavity ($R<\Rph$), the $\lambda$ bound scales up as $\Rph/R$; if the optical continuum is a dust-reddened disk without a photosphere \citep{Madau2026a}, $\Rph$ disappears while the Section~\ref{sec:budget} bounds ($\sigma_{M/R}\leq\sigLamMax{}$~dex and $H/R\simeq1$ or $i>45^\circ$) still hold.
In an orbital reading, the absence of $\Lph^{1/4}\Teff$ scaling in $M/R$ at fixed $\LHa$ is naturally satisfied if $R$ follows the ionizing or recombination luminosity rather than $\Rph$.

\emph{An Eddington or super-Eddington envelope.}
If $\lambda\geq1$ with $R\geq\Rph$, the escape speed is at most $\sqrt{2}\,\vK\,\lambda^{-1/2}$, so the wings beyond $\sqrt{2}\,\vK$ trace unbound outflow or post-emission scattering.
A gravitational core at $\Rph$ must then be broadened by at least $\lambda^{1/2}f_{\rm vir}^{1/2}(\ratLo{}$--$\ratHi{})$: \brdVirLo{}--\brdVirHi{} times for $f_{\rm vir}=1.08$ ($k=\kVir{}$), \brdIsoLo{}--\brdIsoHi{} times for $f_{\rm vir}=0.54$ ($k=\kIso{}$), and \brdRingLo{}--\brdRingHi{} times for $f_{\rm vir}=0.25$ ($k=\kRing{}$) at $\lambda=1$.
Electron scattering at $\tau_e\simeq2$--3 and $T_e\sim10^{4}$~K can maintain lines this broad \citep{Inayoshi2026}, with radiative cooling stabilizing $T_e$ near $10^{4}$~K except at very low metallicity \citep{Sneppen2026}.
Using the exponential wing scale $W\simeq(\rusA{}\,\tau_e+\rusB{})(T_e/10^{4}~{\rm K})^{1/2}$~\kms{} \citep{Rusakov2026}, the sensitivity $D\equiv\partial\log W/\partial\log\tau_e$ is \DTwo{}--\DThree{} across $\tau_e=2$--3.
Our ceiling $\sigint<\FsigUp{}$~dex therefore restricts $\sigma_{\log\tau_e}$ at fixed $\LHa$ to at most \sigTauThree{}--\sigTauTwo{}~dex (\sigTauConvHiLo{}--\sigTauConvHiHi{}~dex when convolved with a \coreLo{}--\coreHi{}~\kms{} core; Appendix~\ref{app:scattering}), or $\sigma_{\log T_e}$ to at most \sigTeMax{}~dex.

That $W(\tau_e)$ calibration injects photons into an external scattering shell; where recombination and Thomson scattering share one ionized zone instead, $\LHa\propto R\,\tau_e^{2}$ \citep{Rusakov2026,Sneppen2026}.
This coupling is consistent with widths that follow $\LHa$ rather than the continuum photosphere (Figure~\ref{fig:photo}c).
Because $\tau_e$ is tied to the recombination emission itself, rising between $\LHa^{1/4}$ (if $R\propto\LHa^{1/2}$) and $\LHa^{1/2}$ (fixed $R$), scattered widths scale as $\LHa^{\shellSlopeLo{}\text{--}\shellSlopeHi{}}$ at the $\tau_e\simeq2$--3 required above (Appendix~\ref{app:scattering}), and only as $\LHa^{\tauPhLo{}}$ at $\tau_e=0.5$, which supplies too little broadening.
This matches our baseline slope (\Fb{}) and that of \BX{} (\Xb{}), but the slopes after correcting the narrow-line split (\FbNarFix{}--\FbNarOIII{}, with 95\% upper ends \FbNarFixhi{}--\FbNarOIIIhi{}) lie at or below the lower edge of this range.
Writing $\LHa=K R\,\tau_e^{2}$ at fixed emissivity/geometry factor $K$ and $T_e$, $\tau_e\propto R^{-1/2}$ also bounds shell-radius variations at fixed $\LHa$ to $\sigma_{\log R}\simeq2\,\sigint/D\lesssim\sigRThree{}\text{--}\sigRTwo{}$~dex across $\tau_e=2$--3 (Appendix~\ref{app:scattering}).

\section{Discussion} \label{sec:discussion}

\subsection{Flattened Thermal Disks, Compact Binaries, and Non-AGN Alternatives} \label{sec:alternatives}

With $\Rph\equiv(\Lph/4\pi\sigma_{\rm SB}\Teff^{4})^{1/2}$ serving as an empirical continuum reference radius, Equation~(\ref{eq:ratio}) requires any bound virial BLR matching ${\rm FWHM}=(\ratLo{}\text{--}\ratHi{})\,\vK(\Rph)$ to satisfy $\lambda R/\Rph=\lamVirLo{}\text{--}\lamIsoHi{}$ for $f_{\rm vir}=1.08\text{--}0.54$ ($\lamRingLo{}\text{--}\lamRingHi{}$ for $f_{\rm vir}=0.25$).
In Section~\ref{sec:readings}, tension with Eddington-limited envelopes arose because a spherical photosphere simultaneously blocks interior sightlines ($R\geq\Rph$, forcing $\lambda\leq\lamVirLo{}\text{--}\lamIsoHi{}$) and assumes near- or super-Eddington radiation support ($\lambda\gtrsim1$).
Flattened cool ($\sim2000$--$5000$~K) thermal-disk models---a gravitationally unstable ($Q\sim1$) outer disk paired with an inner standard disk \citep{ZhangCX2026}, the $T\simeq5000$~K inner rim of a circumbinary disk carved by a compact ($\lesssim10^{3}$ Schwarzschild radii) massive black hole binary \citep{InayoshiChen2026}, or the outer ``Hayashi limit'' zone of an extended self-gravitating disk \citep{ChenYX2026}---decouple these two conditions in either of two regimes.
First, when the cool outer disk or cavity rim surrounds a $10^{7}$--$10^{8}\,\Msun$ black hole or binary without spherical radiation support, its optical luminosity satisfies $\lambda\equiv\Lph/L_{\rm Edd}\ll1$; bound virial gas at $R\simeq(\lamVirLo{}\text{--}\lamIsoHi{})\,\lambda^{-1}\Rph\gtrsim\Rph$ (which can lie near or inside the physical disk or rim radius $R_{\rm disk}\geq\Rph$ when inclined or annular emission has projected area $<4\pi R_{\rm disk}^{2}$) then matches ${\rm FWHM}\simeq(\ratLo{}\text{--}\ratHi{})\,\vK(\Rph)$ and wings beyond $\sqrt{2}\,\vK(\Rph)$.
Second, because a flattened disk does not occult inclined sightlines to smaller radii, interior bound orbits at $R\simeq(\lamVirLo{}\text{--}\lamIsoHi{})\,\lambda^{-1}\Rph<\Rph$ can match the same widths even at $\lambda\sim1$, lifting the spherical sub-Eddington ceiling.
In both regimes, modest dust attenuation ($A_V\lesssim1$~mag) satisfies $\sigma_A\leq\sigDustMax{}$~mag (Section~\ref{sec:budget}), and spatial separation between the cool outer continuum zone and the central ionizing UV source allows broad-\Ha{} widths to track $\LHa$ rather than $\Lph^{1/4}\Teff$.

Even where the spherical $\lambda$ bound is lifted, $\sigint<\FsigUp{}$~dex constrains the line-emitting dynamics.
An outer or circumbinary virial BLR still requires $\sigma_{M/R}\leq\sigLamMax{}$~dex alongside thick kinematics or restricted inclinations (Equation~\ref{eq:budget}), unless $\log f_{\rm vir}$ and $\log(M/R)$ positively covary.
In compact-binary models \citep{InayoshiChen2026}, any broad-\Ha{} component emitted from mini-disks or cavity streams modulated by the binary orbital speed $v_{\rm bin}\sim(GM_{\rm tot}/a_{\rm bin})^{1/2}$ must produce an intrinsic width dispersion below $\FsigUp{}$~dex across binary separations, mass ratios, orbital phases, and inclinations at fixed $\LHa$ to remain consistent with the data; likewise, an unscattered bulk wind launched near $\Rph$ at $\lambda\geq1$ would require multi-escape terminal speeds scaling with $\LHa$ rather than $\vK(\Rph)$ while keeping object-to-object scatter below $\FsigUp{}$~dex.

Non-AGN scenarios face distinct empirical constraints.
The unresolved [\ion{O}{3}]~$\lambda5007$ widths (at the \wnarLo{}~\kms{} bound in \NOIIIUnres{} of \NOIIIMeas{} measured LRDs; Section~\ref{sec:variants}) do not support attributing the median \WLRDmed{}~\kms{} broad \Ha{} emission directly to the same low-density, galaxy-scale kinematic component \citep[cf.][]{Baggen2024} or kiloparsec-scale ionized outflows, although narrow [\ion{O}{3}] alone neither determines the density or spatial scale of the broad-\Ha{} gas nor excludes more complex host environments.
Recurring cluster tidal disruption events sampled across fallback phases \citep{Bellovary2025} must similarly keep width dispersion below $\FsigUp{}$~dex at fixed $\LHa$ alongside weak continuum variability, while an Eddington-limited $\Teff\simeq5000$~K supermassive star \citep{NandalLoeb2026} obeys Equation~(\ref{eq:vk}) and shares the Section~\ref{sec:readings} requirement of uniform non-gravitational broadening.

\subsection{Consequences for Virial Black Hole Masses and Demographics} \label{sec:mvir}

Regardless of which reading holds, single-epoch virial estimators compute $\Mvir\propto\LHa^{\virL{}}\,{\rm FWHM}^{\virW{}}$ \citep[][with $f_{\rm vir}=1.08$, $k=\kVir{}$ from \citealt{ReinesVolonteri2015}]{Reines2013}.
At fixed observed $\LHa$, our limit $\sigint<\FsigUp{}$~dex restricts the intrinsic scatter of $\Mvir$ to less than \sigMvir{}~dex (\virW{}\,$\sigint$ rounded up; \sigMvirMed{}~dex at the formal-error median, $<\sigMvirRep{}$~dex with the same-grating repeat floor, and $<\sigMvirCons{}$~dex at the conservative ceiling).
The virial masses tabulated by \BX{} are consistent with this: including measurement error, their residuals around a power law in luminosity have a standard deviation of \XresM{}~dex for LRDs versus \XresMo{}~dex for comparison sources, although the latter value also reflects heterogeneous mass estimators.
For LRDs, single-epoch virial masses, together with mass functions or \Ha{}-based Eddington ratios built from them, are nearly deterministic restatements of line luminosity, whereas the true central mass differs by one to three orders of magnitude between the orbital and envelope readings (Table~\ref{tab:readings}).

This near-degeneracy between $\Mvir$ and $\LHa$ has direct consequences for high-redshift black hole demographics.
First, because $\Mvir$ follows $\LHa^{\virL{}+\virW{}b}$ with less than \sigMvir{}~dex of intrinsic dispersion, within a given single-epoch virial calibration, virial black hole mass functions constructed for LRDs \citep[e.g.,][]{Matthee2024,Taylor2025} are essentially a power-law coordinate transformation of the broad-\Ha{} luminosity function rather than an independent measure of the mass distribution: unlike lower-redshift quasars, where substantial width dispersion broadens the high-mass tail via Eddington-like upscatter \citep{Shen2008,Croom2011}, intrinsic width scatter in LRDs induces far less upscatter at a given mass-function slope, while leaving the overall mass normalization uncertain by $1$--$3$~dex across models.
Second, at fixed host stellar mass or upper limit, object-to-object scatter in virial $M_{\rm BH}/M_*$ ratios across LRDs chiefly reflects variations in $\LHa/M_*$ rather than independent dynamical mass spread.
Pinning down the absolute mass zero point between the standard-virial branch ($10^{7}$--$10^{8}\,\Msun$, at $R\gg\Rph$) and lighter envelope or interior-orbit models ($10^{4}$--$10^{7}\,\Msun$) therefore governs whether LRDs require rapid early assembly of massive black holes \citep{InayoshiIchikawa2024,Inayoshi2025} operating at modest bolometric corrections consistent with the So{\l}tan radiative budget \citep{Soltan1982,Greene2026,InayoshiChen2026}, or lower-mass seeds growing inside super-Eddington envelopes near local $M_{\rm BH}$--$M_*$ scaling relations \citep{BegelmanDexter2025,Rusakov2026,Sun2026,Curtis2026}.

\subsection{Caveats and Observational Prospects} \label{sec:prospects}

Several scope limitations of the present sample should be kept in mind.
Our uniformly refitted JWST sample contains \NLRD{} LRDs at $z=\zLRDmin{}$--$\zLRDmax{}$ spanning $\log(\LHa/\ergs)=\LLRDmin{}$--$\LLRDmax{}$, with only \NBright{} luminous objects near $10^{\LBright{}}$~\ergs{}, leaving the width--luminosity slope sensitive to bright-end leverage and to how the line core is partitioned between narrow and broad components ($b=\FbNarFix{}$--\Fb{}; Sections~\ref{sec:relation} and \ref{sec:variants}).
Likewise, the subsamples with published spectroscopic $(\Lph,\Teff)$ measurements comprise \NPE{} to \NPA{} objects, and the flat partial dependence on $\Lph$ at fixed $\LHa$ rests heavily on the leverage of RUBIES-UDS-29813 (Section~\ref{sec:photosphere}).
Finally, our baseline regressions use broad $\LHa$ uncorrected for dust; differential attenuation enters the scatter budget through $(0.4\,b\,\sigma_A)^2$ (Section~\ref{sec:budget}), whereas dereddening via the Balmer decrement in \BX{} leaves the small LRD scatter intact ($\Xsig{}$~dex; Section~\ref{sec:relation}), with the caveat that elevated broad-line \Ha/\Hb{} ratios in dense gas ($n_{\rm H}\gtrsim10^{9}$~cm$^{-3}$) also reflect collisional excitation and radiative transfer rather than dust alone \citep{Chang2025,ChenCH2026}.
Our constraints likewise apply to spectroscopically confirmed broad-\Ha{} LRDs rather than continuum-selected candidates lacking broad Balmer emission.

Four observational tests can break the remaining degeneracy between sub-Eddington or flattened-disk bound orbits and Eddington-limited scattering envelopes:
(i) expanding high-resolution ($R\gtrsim2700$) NIRSpec grating samples at the bright end ($\LHa\gtrsim10^{43}$~\ergs{}), with [\ion{O}{3}]-anchored narrow-line subtraction, to test whether widths follow the steeper scaling (${\rm FWHM}\propto\LHa^{\shellSlopeLo{}\text{--}\shellSlopeHi{}}$) required by cospatial recombination and Thomson scattering at $\tau_e\simeq2$--3 or remain nearly flat ($b\lesssim0.1$) as permitted by bound orbits;
(ii) obtaining uniform continuum plus grating spectroscopy across a wider dynamic range in $\Lph/\LHa$ to test photospheric $\Lph^{1/4}\Teff$ scaling without reliance on single outlier objects;
(iii) measuring high-S/N multi-transition Balmer and Paschen profiles \citep{Brazzini2025} alongside velocity-resolved decrements \citep{ChenCH2026} to separate radial density stratification in a virial BLR \citep{Madau2026b,Scholtz2026} from wavelength-independent Thomson wings \citep{Rusakov2026}, selective resonance/Raman wings \citep{Chang2025}, or orbital velocity shifts in compact binaries \citep{InayoshiChen2026}; and
(iv) securing additional spatially resolved dynamical masses in strongly lensed LRDs \citep{Juodzbalis2026} alongside multi-year spectroscopic monitoring for reverberation lags or radial-velocity drifts, directly separating $10^{7}$--$10^{8}\,\Msun$ potentials from $10^{4}$--$10^{7}\,\Msun$ envelopes.

\section{Summary and Conclusions} \label{sec:summary}

We have measured the intrinsic scatter and photospheric velocity scaling of broad \Ha{} emission lines in high-redshift little red dots by uniformly refitting public JWST/NIRSpec grating spectra alongside lower-redshift SDSS quasars. Our main findings are as follows:

\begin{enumerate}
\item Broad \Ha{} widths in \NLRD{} LRDs have $\sigint=\Fsig{}$~dex at fixed $\LHa$ under formal errors ($<\FsigUp{}$~dex at 95\% in our baseline fits, $\lesssim\FsigCons{}$~dex across five of six uniform width measures and analysis variants, and $<\reAllUp{}$--$\reMultUp{}$~dex from duplicate spectra depending on the error model), and are consistent with a constant \WConstMed{}~\kms{} below the \NBright{} brightest LRDs. Other JWST broad-line sources scatter by \sigTwoLo{}--\sigTwoHi{}~dex, and identically refitted SDSS quasars are \sdssRatSame{} times more scattered (\Ssg{}~dex) and \sdssWoff{}~dex wider at the median LRD luminosity.
\item Relative to the Eddington-photosphere Keplerian speed $\vK(\Rph)\propto\Lph^{1/4}\Teff$ evaluated from each LRD's measured continuum luminosity and temperature, median Gaussian widths are \ratLo{}--\ratHi{}\,$\vK$ and \fEscNpLo{}--\fEscNpHi{}\% of the broad-\Ha{} flux lies beyond $\sqrt{2}\,\vK$, yet at fixed $\LHa$ we detect no dependence on $\Lph$ (partial slope \bpJA{}, \bpJAlo{} to \bpJAhi{}), a null result that becomes consistent with photospheric scaling when its most leveraged source is removed.
\item Competing physical models face distinct quantitative requirements (Table~\ref{tab:readings}): bound orbits outside a spherical photosphere need $\lambda\leq\lamVirLo{}\text{--}\lamIsoHi{}$ for standard or isotropic geometries ($f_{\rm vir}=1.08\text{--}0.54$; $\leq\lamRingHi{}$ for an edge-on ring, $f_{\rm vir}=0.25$) alongside $\sigma_{M/R}\leq\sigLamMax{}$~dex and thick kinematics or restricted inclinations ($H/R\simeq1$ or $i>45^\circ$) under independent orientation and $M/R$ variations, with a photoionized BLR ($R$ tracking $\LHa$) naturally matching the flat $\Lph$ residuals; flattened cool-continuum models (unstable outer, circumbinary, or self-gravitating disks) avoid spherical-envelope tension---either satisfying the $\lambda$ bound at $\lambda\ll1$ with $R\simeq(\lamVirLo{}\text{--}\lamIsoHi{})\,\lambda^{-1}\Rph\gtrsim\Rph$ or lifting it via unobscured interior orbits at $R<\Rph$---while still requiring net orbital, inclination, and binary-modulation scatter below $\FsigUp{}$~dex at fixed $\LHa$; and an Eddington or super-Eddington spherical envelope ($\lambda\geq1$) needs non-gravitational broadening by $f_{\rm vir}^{1/2}(\ratLo{}\text{--}\ratHi{})$ at $\lambda=1$ ($\tau_e\simeq2$--3 at $10^{4}$~K) and $\sigma_{\log\tau_e}\leq\sigTauThree{}\text{--}\sigTauTwo{}$~dex under isothermal-shell scaling, with cospatial recombination and scattering ($\LHa\propto R\,\tau_e^{2}$) providing a compatible $\LHa$ coupling.
\item Within a given single-epoch virial calibration and independent of the true continuum or line-broadening mechanism, virial mass estimates of LRDs scatter intrinsically by less than \sigMvir{}~dex at fixed observed \Ha{} luminosity (\sigMvirMed{}~dex at the formal median, $<\sigMvirCons{}$~dex at the conservative ceiling), rendering single-epoch virial masses and virial black hole mass functions nearly deterministic transformations of line luminosity while the true mass zero point remains separated by $1$--$3$~dex across models.
\end{enumerate}

\begin{acknowledgments}
The author is grateful to his parents for their constant support. This research received no external grant funding.

Some of the data products presented herein were retrieved from the Dawn JWST Archive (DJA). DJA is an initiative of the Cosmic Dawn Center (DAWN), which is funded by the Danish National Research Foundation under grant DNRF140.
This work is based on observations made with the NASA/ESA/CSA James Webb Space Telescope. The reduced spectra used here were retrieved from the DAWN JWST Archive; the underlying JWST observations are archived in the Mikulski Archive for Space Telescopes at the Space Telescope Science Institute, which is operated by the Association of Universities for Research in Astronomy, Inc., under NASA contract NAS5-03127 for JWST. These observations are associated with programs \#1345 (PI S. Finkelstein), \#3567 (PI F. Valentino), \#4106 (PI E. Nelson), \#4233 (PI A. de Graaff), \#5718 (PI D. D. Kocevski), \#8047 (PI B. Wang), \#8410 (PI R. Simons), and \#9214 (PI C. Mason). We acknowledge the teams of programs \#1345, \#4233, \#8410, and \#9214 for developing their observing programs with zero-exclusive-access periods.

Funding for the SDSS and SDSS-II has been provided by the Alfred P. Sloan Foundation, the Participating Institutions, the National Science Foundation, the U.S. Department of Energy, the National Aeronautics and Space Administration, the Japanese Monbukagakusho, the Max Planck Society, and the Higher Education Funding Council for England. The SDSS Web Site is http://www.sdss.org/. The SDSS is managed by the Astrophysical Research Consortium for the Participating Institutions.

The Participating Institutions are the American Museum of Natural History, Astrophysical Institute Potsdam, University of Basel, University of Cambridge, Case Western Reserve University, University of Chicago, Drexel University, Fermilab, the Institute for Advanced Study, the Japan Participation Group, Johns Hopkins University, the Joint Institute for Nuclear Astrophysics, the Kavli Institute for Particle Astrophysics and Cosmology, the Korean Scientist Group, the Chinese Academy of Sciences (LAMOST), Los Alamos National Laboratory, the Max-Planck-Institute for Astronomy (MPIA), the Max-Planck-Institute for Astrophysics (MPA), New Mexico State University, Ohio State University, University of Pittsburgh, University of Portsmouth, Princeton University, the United States Naval Observatory, and the University of Washington.

AI disclosure: Gemini (Google), Claude (Anthropic), Codex (OpenAI), and DeepSeek were used under the author's direction to audit physical and statistical arguments, write and execute analysis and Matplotlib figure-generation code, cross-check literature citations, and refine prose from the author's drafts. All text, numbers, figures, and references were verified by the author, who bears full responsibility for the content.
\end{acknowledgments}

\begin{contribution}
H.W. conceived and designed the study, performed the data analysis and modeling, and wrote the manuscript.
\end{contribution}

\facilities{JWST (NIRSpec), Sloan}

\software{\texttt{msaexp} \citep{msaexp}, NumPy \citep{harris2020array}, SciPy \citep{2020SciPy-NMeth}, Matplotlib \citep{Hunter:2007}, Astropy \citep{astropy:2013,astropy:2018,astropy:2022}}

\clearpage
\restartappendixnumbering
\appendix
\twocolumngrid

\section{Photospheric Subsamples and Scaling Regressions} \label{app:scaling}

The atmosphere fits of \citet{Curtis2026} match \NPAall{} LRDs in \BX{} and \NPBall{} in our sample.
Five have $\log\Lph$ posteriors \sxDropMin{}--\sxDropMax{}~dex wide (versus $\leq\sxKeepMax{}$~dex for the rest), which carry negligible leverage and bias conditional likelihoods of the form of Equation~(\ref{eq:var}) through $\ln V_i$.
Excluding those \NPdrop{} leaves Sample~A (\NPA{} sources, \BX{} widths) and Sample~B (\NPB{} sources, our widths), while retaining them in a hierarchical Gaussian-prior regression shifts the $\Lph$ exponent by less than \hierDb{}.
Modified-blackbody parameters are published for \NdGbb{} of the \NdGall{} LRDs of \citet{deGraaff2026} ($z<4.5$); \NdGfit{} have grating \Ha{} coverage, \NPE{} pass Section~\ref{sec:sample} (Sample~E), and \BX{} adds \NPCbx{} more (Sample~C, \NPC{} sources).
Unpublished $\Lph$--$\Teff$ covariances are set to zero.

Fitting $\log{\rm FWHM}=a+b\log\Lph+c\log\Teff$ yields $b=\bPA{}$ (\bPAlo{}--\bPAhi{}), $c=\cPA{}$ (\cPAlo{} to \cPAhi{}) on Sample~A; $b=\bPB{}$ (\bPBlo{}--\bPBhi{}), $c=\cPB{}$ (\cPBlo{} to \cPBhi{}) on Sample~B; and $b=\bPC{}$ (\bPClo{}--\bPChi{}), $c=\cPC{}$ (\cPClo{} to \cPChi{}) on Sample~C (Figure~\ref{fig:photo}b).
A width governed by $\LHa$ alone projects onto $b\approx\lineB{}$ and $c=\lineCLo{}$--\lineCHi{} in Samples~A and B via $\Lph$--$\LHa$ correlation, matching the observed posteriors; fixed-scaling scatters around $\Lph^{1/4}\Teff$ are \sqPA{}, \sqPB{}, and \sqPC{}~dex (\sqPElo{}--\sqPEhi{}~dex on Sample~E).
In joint regressions on $\log\LHa$ and $\log\Lph$ using $\LHa$ from our fits (Figure~\ref{fig:photo}c), simulated populations obeying Equation~(\ref{eq:ratio}) at fixed $\lambda R/\Rph$ yield median $\Lph$ partial slopes of \simMedJA{}, \simMedJB{}, and \simMedJE{} on Samples~A (\NJA{} objects with our $\LHa$), B (\NJB{}), and E (\NJE{}).
Sample~E returns \bpJE{} (\bpJElo{} to \bpJEhi{}; \powLo{}--\powHi{}\% power at its median scatter, \powTen{}\% at \simA{}~dex), whereas Samples~A and B return \bpJA{} and \bpJB{}, matched by only \simPctJA{}\% and \simPctJB{}\% of photospheric realizations.

\section{Injection--Recovery Suite} \label{app:injections}

Table~\ref{tab:injections} summarizes the \NReal{}-realization injection--recovery experiments referenced in Section~\ref{sec:injtests}.
Synthetic broad lines are injected onto the pixel grids, noise vectors, and grating LSFs of the \NLRD{} surviving LRD spectra, refitted with the pipeline of Section~\ref{sec:fit}, and filtered by the quality cuts of Section~\ref{sec:sample}.
The bottom block of Table~\ref{tab:injections} extends the suite to all \NParentInBand{} in-band parent LRD spectra, using fixed narrow-line baselines with refitted broad $\LHa$ where ${\rm S/N}\geq\cutSNthree{}$ and published broad $\LHa$ or the narrow-line flux scale where ${\rm S/N}<\cutSNthree{}$ among the \NParentFail{} failed spectra.
Fitting mismatched Lorentzian or exponential profiles to Gaussian lines inflates $\sigint$ to \recGL{} and \recGE{}~dex at $\sigma_{\rm parent}=\simInj{}$~dex, whereas the single-Gaussian FWHM recovers \recGG{}, \recEG{}, and \recLG{}~dex on Gaussian, exponential, and Lorentzian injections without compressing true scatter.
Shifting parent exponential widths upward by \wideInjShift{}~dex allows the ${\rm S/N}\geq\cutSN{}$ cut to clip the widest tail at $\sigma_{\rm parent}=\simA{}$~dex (lowering the median recovered $\sigint$ from \recTenE{} to \recTenWide{}~dex, with \hitTenWide{} of \NReal{} trials $\leq\Fsig{}$~dex), at the cost of shifting the surviving sample \wideSurvShift{}~dex wider than observed.
Finally, residuals of the observed LRD relation show no correlation with $\log({\rm S/N})$ ($r=\rResSN{}$, $p=\pResSN{}$), consistent with the null-slope ($b=0$) rows of Table~\ref{tab:injections}.
Two simplifications limit these suites: injected lines carry no absorption, although \NAbsLRD{} LRDs (including the \NBright{} most luminous) are fitted with absorption, and the injected noise equals the formal pixel errors, so the reduced-$\chi^2$ rule never rejects a trial and the unmodeled variance seen in repeat spectra (Section~\ref{sec:repeats}) is not represented.

\input{tables/TableB1_injections}

\section{Sensitivity of Single-Gaussian Widths to Electron-Scattering Optical Depth} \label{app:scattering}

For pure exponential wings of $e$-folding scale $W=(\rusA{}\,\tau_e+\rusB{})(T_e/10^{4}~{\rm K})^{1/2}$~\kms{} \citep{Rusakov2026}, $D\equiv\partial\log W/\partial\log\tau_e$ equals \DHalf{}, \DOne{}, \DTwo{}, and \DThree{} at $\tau_e=0.5$, 1, 2, and 3.
Evaluated at $\sigint=\FsigUp{}$~dex, $\sigma_{\log\tau_e}\leq\sigint/D$ gives \sigTauHalf{}, \sigTauOne{}, \sigTauTwo{}, and \sigTauThree{}~dex, and the shell-radius budget $\sigma_{\log R}\simeq2\,\sigint/D$ at fixed $K$ and $T_e$ gives \sigRHalf{}, \sigROne{}, \sigRTwo{}, and \sigRThree{}~dex.
When a \coreLo{}--\coreHi{}~\kms{} Gaussian core is convolved with that exponential kernel and fitted with a single Gaussian, core dilution lowers $D_{\rm eff}\equiv\partial\log{\rm FWHM}/\partial\log\tau_e$ to \DconvHalfLo{}--\DconvHalfHi{} at $\tau_e=0.5$ ($\sigma_{\log\tau_e}\leq\sigTauConvHalfLo{}$--\sigTauConvHalfHi{}~dex) and \DconvHiLo{}--\DconvHiHi{} across $\tau_e=2$--3 ($\sigma_{\log\tau_e}\leq\sigTauConvHiLo{}$--\sigTauConvHiHi{}~dex; $\sigma_{\log R}\lesssim\sigRConvHiLo{}$--\sigRConvHiHi{}~dex).
Retaining an unscattered transmitted core $e^{-\tau_e}$ alongside the scattered $(1-e^{-\tau_e})$ component keeps the fitted Gaussian FWHM within \rtHalfBoostMax{}\% of the intrinsic core at $\tau_e=0.5$ while steepening $D_{\rm eff}$ across $\tau_e=2$--3 to \DrtHiLo{}--\DrtHiHi{}, which tightens the $\tau_e$ and $R$ scatter bounds relative to the pure-wing values.
For $\tau_e$ rising as $\LHa^{1/4}$ to $\LHa^{1/2}$ (Section~\ref{sec:readings}), the predicted width--luminosity slope is $D/4$ to $D/2$: \shellPureLo{}--\shellSlopeHi{} for pure wings and \shellSlopeLo{}--\shellConvHi{} with core dilution across $\tau_e=2$--3, and steeper with a transmitted core.

\bibliography{refs,refs_ads_20261113}
\bibliographystyle{aasjournalv7}

\end{document}

%% file: numbers.tex
\newcommand{\NLRD}{29}
\newcommand{\NOther}{15}
\newcommand{\NFrozen}{44}
\newcommand{\NSources}{115}
\newcommand{\NSpectra}{204}
\newcommand{\NSpecHa}{177}
\newcommand{\NAbsLRD}{five}
\newcommand{\NAbsOther}{four}
\newcommand{\cutSN}{5}
\newcommand{\cutSNthree}{3}
\newcommand{\cutBIC}{6}
\newcommand{\cutChi}{2}
\newcommand{\HvidingCut}{1000}
\newcommand{\wnarDef}{150}
\newcommand{\passShift}{0.10}
\newcommand{\rOutRph}{100}
\newcommand{\matchRadius}{0\farcs5}
\newcommand{\zLRDmin}{3.36}
\newcommand{\zLRDmax}{7.15}
\newcommand{\SNLRDmed}{13.7}
\newcommand{\SNOthmed}{8.2}
\newcommand{\zOthmin}{3.60}
\newcommand{\zOthmax}{6.04}
\newcommand{\LLRDmin}{41.82}
\newcommand{\LLRDmed}{42.34}
\newcommand{\LLRDmax}{43.55}
\newcommand{\WLRDmin}{945}
\newcommand{\WLRDmax}{3185}
\newcommand{\WLRDmed}{1777}

\newcommand{\NBICwing}{25}
\newcommand{\NBICsix}{16}
\newcommand{\NBIChiSN}{nine}
\newcommand{\NBICsixOth}{four}
\newcommand{\NCoreWing}{seven}

\newcommand{\Fsig}{0.054}
\newcommand{\Fsiglo}{0.038}
\newcommand{\Fsighi}{0.080}
\newcommand{\FsigUp}{0.075}
\newcommand{\Fb}{0.17}
\newcommand{\Fblo}{0.12}
\newcommand{\Fbhi}{0.22}
\newcommand{\Fa}{3.346}
\newcommand{\Fsigo}{0.154}
\newcommand{\Fsigolo}{0.104}
\newcommand{\Fsigohi}{0.252}
\newcommand{\Fdsig}{\ensuremath{-}0.099}
\newcommand{\Frms}{0.067}
\newcommand{\NNoTop}{26}
\newcommand{\FbNoTop}{0.09}
\newcommand{\FbNoToplo}{0.01}
\newcommand{\FbNoTophi}{0.19}
\newcommand{\FsigNoTop}{0.040}
\newcommand{\FsigNoToplo}{0.020}
\newcommand{\FsigNoTophi}{0.069}
\newcommand{\NBright}{three}
\newcommand{\LBright}{43.5}
\newcommand{\LRestMax}{42.9}
\newcommand{\NNoAbs}{24}
\newcommand{\FsigNoAbs}{0.043}
\newcommand{\FsigNoAbslo}{0.022}
\newcommand{\FsigNoAbshi}{0.074}
\newcommand{\FbNoAbs}{0.11}
\newcommand{\FbNoAbslo}{0.02}
\newcommand{\FbNoAbshi}{0.22}
\newcommand{\FdsigNoAbs}{\ensuremath{-}0.111}
\newcommand{\FdsigNoAbslo}{\ensuremath{-}0.211}
\newcommand{\FdsigNoAbshi}{\ensuremath{-}0.051}
\newcommand{\FdsigDroplo}{\ensuremath{-}0.119}
\newcommand{\FdsigDrophi}{0.002}
\newcommand{\WOutA}{3612}
\newcommand{\WOutB}{2915}
\newcommand{\FsigLsf}{0.069}
\newcommand{\FsigLsfUp}{0.094}
\newcommand{\NNarBoundLRD}{two}
\newcommand{\WNarBoundBefore}{3150}
\newcommand{\WNarBoundAfter}{2400}
\newcommand{\NLRDNarCut}{27}
\newcommand{\FbNarCut}{0.11}
\newcommand{\FbNarCutlo}{0.05}
\newcommand{\FbNarCuthi}{0.17}
\newcommand{\FsigNarCut}{0.040}
\newcommand{\FsigNarCutlo}{0.023}
\newcommand{\FsigNarCuthi}{0.067}
\newcommand{\FsigNarCutUp}{0.062}
\newcommand{\FdsigNarCut}{\ensuremath{-}0.113}
\newcommand{\FdsigNarCutlo}{\ensuremath{-}0.213}
\newcommand{\FdsigNarCuthi}{\ensuremath{-}0.057}
\newcommand{\FbNarFix}{0.09}
\newcommand{\FbNarFixlo}{0.04}
\newcommand{\FbNarFixhi}{0.13}
\newcommand{\FsigNarFix}{0.046}
\newcommand{\FsigNarFixlo}{0.030}
\newcommand{\FsigNarFixhi}{0.069}
\newcommand{\FsigNarFixUp}{0.064}
\newcommand{\FsigoNarFix}{0.134}
\newcommand{\FsigoNarFixlo}{0.090}
\newcommand{\FsigoNarFixhi}{0.221}
\newcommand{\FdsigNarFix}{\ensuremath{-}0.088}
\newcommand{\FdsigNarFixlo}{\ensuremath{-}0.176}
\newcommand{\FdsigNarFixhi}{\ensuremath{-}0.038}
\newcommand{\FbNarOIII}{0.11}
\newcommand{\FbNarOIIIlo}{0.05}
\newcommand{\FbNarOIIIhi}{0.17}
\newcommand{\FsigNarOIII}{0.072}
\newcommand{\FsigNarOIIIlo}{0.052}
\newcommand{\FsigNarOIIIhi}{0.102}
\newcommand{\FsigNarOIIIUp}{0.097}
\newcommand{\FsigoNarOIII}{0.150}
\newcommand{\FsigoNarOIIIlo}{0.099}
\newcommand{\FsigoNarOIIIhi}{0.248}
\newcommand{\FdsigNarOIII}{\ensuremath{-}0.077}
\newcommand{\FdsigNarOIIIlo}{\ensuremath{-}0.178}
\newcommand{\FdsigNarOIIIhi}{\ensuremath{-}0.018}
\newcommand{\FsigNarMin}{0.070}
\newcommand{\FsigNarMinUp}{0.094}
\newcommand{\NOIIIUnres}{21}
\newcommand{\FsigCons}{0.10}
\newcommand{\narCeilLoosenPct}{35}
\newcommand{\FsigOIIIqc}{0.075}
\newcommand{\FsigOIIIqcUp}{0.102}
\newcommand{\NOIIIqc}{27}
\newcommand{\FsigWinWide}{0.083}
\newcommand{\FsigWinWidelo}{0.059}
\newcommand{\FsigWinWidehi}{0.119}
\newcommand{\winWider}{8000}
\newcommand{\NNarThirty}{17}
\newcommand{\NOIIIMeas}{23}
\newcommand{\dWNarRms}{0.06}
\newcommand{\dWNarMax}{0.21}
\newcommand{\NFailSNthreeWeak}{seven}
\newcommand{\FsigConst}{0.035}
\newcommand{\FsigConstlo}{0.015}
\newcommand{\FsigConsthi}{0.067}
\newcommand{\WConstMed}{1760}
\newcommand{\LConstMin}{41.8}
\newcommand{\LConstMax}{42.9}
\newcommand{\bpJAdrop}{0.02}
\newcommand{\bpJAdroplo}{\ensuremath{-}0.16}
\newcommand{\bpJAdrophi}{0.22}
\newcommand{\bpJBdrop}{\ensuremath{-}0.04}
\newcommand{\bpJBdroplo}{\ensuremath{-}0.33}
\newcommand{\bpJBdrophi}{0.24}
\newcommand{\dLphOut}{0.5}
\newcommand{\dsigInflLo}{\ensuremath{-}0.12}
\newcommand{\dsigInflHi}{\ensuremath{-}0.08}
\newcommand{\inflLo}{1.5}
\newcommand{\inflHi}{2.8}
\newcommand{\inflFlrLo}{0.044}
\newcommand{\inflFlrHi}{0.08}
\newcommand{\NGHobj}{seven}
\newcommand{\dWGH}{\ensuremath{-}0.06}
\newcommand{\offGH}{0.054}
\newcommand{\offGHerr}{0.022}
\newcommand{\dWPrim}{\ensuremath{-}0.056}
\newcommand{\NGM}{22}
\newcommand{\FsigGM}{0.048}
\newcommand{\FsigGMUp}{0.072}
\newcommand{\FsigSwap}{0.052}
\newcommand{\FsigSwapUp}{0.072}
\newcommand{\FsigLowSNUp}{0.083}
\newcommand{\SsgNoChi}{0.126}
\newcommand{\NSsgChi}{31}
\newcommand{\WSdssSNfail}{7090}
\newcommand{\shellSlopeLo}{0.14}
\newcommand{\shellSlopeHi}{0.39}
\newcommand{\shellPureLo}{0.17}
\newcommand{\shellConvHi}{0.37}
\newcommand{\sepPair}{35}
\newcommand{\QSOdBICfree}{\ensuremath{-}2.6}
\newcommand{\reMultUp}{0.062}
\newcommand{\reMultScale}{2.0}
\newcommand{\reAddFloor}{0.053}
\newcommand{\NParentClean}{50}
\newcommand{\NParentInBand}{49}
\newcommand{\NParentFail}{20}
\newcommand{\NFailSNthree}{nine}
\newcommand{\NLRDSNthree}{38}
\newcommand{\FbSNthree}{0.17}
\newcommand{\FbSNthreelo}{0.12}
\newcommand{\FbSNthreehi}{0.22}
\newcommand{\FsigSNthree}{0.053}
\newcommand{\FsigSNthreelo}{0.037}
\newcommand{\FsigSNthreehi}{0.077}
\newcommand{\FsigSNthreeUp}{0.072}
\newcommand{\NFailPosUDS}{three}
\newcommand{\resFailPosLo}{0.09}
\newcommand{\resFailPosHi}{0.16}
\newcommand{\failPosSNlo}{4.4}
\newcommand{\failPosSNhi}{5.3}
\newcommand{\NPairsBest}{14}
\newcommand{\NPairsAll}{22}
\newcommand{\NPairsSameG}{seven}
\newcommand{\rmsDupBest}{2.81}
\newcommand{\rmsDupAll}{2.51}
\newcommand{\rmsDupSameG}{1.77}
\newcommand{\flrDupAll}{0.060}
\newcommand{\flrDupSameG}{0.044}
\newcommand{\FsigFlrSame}{0.027}
\newcommand{\FsigFlrSameUp}{0.057}
\newcommand{\FsigFlrAll}{0.018}
\newcommand{\FsigFlrAllUp}{0.049}
\newcommand{\NSNfif}{12}
\newcommand{\rmsSNfif}{0.056}
\newcommand{\NPairsBigObj}{10}
\newcommand{\flrSliceUpLo}{0.049}
\newcommand{\flrSliceUpHi}{0.051}
\newcommand{\NSpecPassLRD}{43}
\newcommand{\reAllUp}{0.049}
\newcommand{\reDropBigUp}{0.057}

\newcommand{\DsigTg}{0.118}
\newcommand{\DsigTglo}{0.082}
\newcommand{\DsigTghi}{0.176}
\newcommand{\sigFiveLo}{0.051}
\newcommand{\sigFiveHi}{0.076}

\newcommand{\simInj}{0.05}
\newcommand{\NReal}{100}
\newcommand{\recGG}{0.052}
\newcommand{\recEG}{0.055}
\newcommand{\recGL}{0.14}
\newcommand{\recGE}{0.10}
\newcommand{\nullDen}{200}
\newcommand{\simA}{0.10}
\newcommand{\recTenE}{0.094}
\newcommand{\wideInjShift}{0.32}
\newcommand{\recTenWide}{0.074}
\newcommand{\hitTenWide}{six}
\newcommand{\wideSurvShift}{0.19}
\newcommand{\recParFiveG}{0.050}
\newcommand{\recParFiveGlo}{0.042}
\newcommand{\recParFiveGhi}{0.060}
\newcommand{\bParFiveG}{0.17}
\newcommand{\bParFiveGlo}{0.14}
\newcommand{\bParFiveGhi}{0.19}
\newcommand{\hitParFiveGnum}{62/100}
\newcommand{\recParFiveE}{0.062}
\newcommand{\recParFiveElo}{0.050}
\newcommand{\recParFiveEhi}{0.073}
\newcommand{\bParFiveE}{0.13}
\newcommand{\bParFiveElo}{0.10}
\newcommand{\bParFiveEhi}{0.16}
\newcommand{\hitParFiveEnum}{27/100}
\newcommand{\recParTenG}{0.101}
\newcommand{\recParTenGlo}{0.085}
\newcommand{\recParTenGhi}{0.120}
\newcommand{\bParTenG}{0.18}
\newcommand{\bParTenGlo}{0.13}
\newcommand{\bParTenGhi}{0.21}
\newcommand{\hitParTenGnum}{0/100}
\newcommand{\hitParTenG}{none}
\newcommand{\recParTenE}{0.092}
\newcommand{\recParTenElo}{0.074}
\newcommand{\recParTenEhi}{0.114}
\newcommand{\bParTenE}{0.13}
\newcommand{\bParTenElo}{0.09}
\newcommand{\bParTenEhi}{0.17}
\newcommand{\hitParTenEnum}{2/100}
\newcommand{\hitParTenE}{two}
\newcommand{\passRatLo}{0.99}
\newcommand{\passRatHi}{1.04}
\newcommand{\NParSelG}{29}
\newcommand{\NParSelGlo}{28}
\newcommand{\NParSelGhi}{31}
\newcommand{\NParSelE}{27}
\newcommand{\NParSelElo}{25}
\newcommand{\NParSelEhi}{29}
\newcommand{\hitParOwnTenGnum}{0/100}
\newcommand{\hitParOwnTenEnum}{1/100}

\newcommand{\NX}{227}
\newcommand{\NXLRD}{48}
\newcommand{\NXCtrl}{97}
\newcommand{\NXOvRec}{29}
\newcommand{\XWcut}{1000}
\newcommand{\XzCut}{3}
\newcommand{\XLmin}{43.7}
\newcommand{\XLmax}{45.2}
\newcommand{\NXLRDuniq}{45}
\newcommand{\XsigDedup}{0.060}
\newcommand{\XsigDeduplo}{0.046}
\newcommand{\XsigDeduphi}{0.080}
\newcommand{\Xsig}{0.058}
\newcommand{\Xsiglo}{0.045}
\newcommand{\Xsighi}{0.077}
\newcommand{\Xb}{0.25}
\newcommand{\Xblo}{0.20}
\newcommand{\Xbhi}{0.31}
\newcommand{\Xsigo}{0.144}
\newcommand{\Xsigolo}{0.122}
\newcommand{\Xsigohi}{0.172}
\newcommand{\Xdsig}{\ensuremath{-}0.085}
\newcommand{\NXNoOv}{28}
\newcommand{\XsigNoOv}{0.075}
\newcommand{\NXPairs}{36}
\newcommand{\XdMatch}{\ensuremath{-}0.065}
\newcommand{\XresM}{0.17}
\newcommand{\XresMo}{0.42}

\newcommand{\NSDSS}{3755}
\newcommand{\Ssig}{0.208}
\newcommand{\zSDSSmin}{0.06}
\newcommand{\zSDSSmax}{0.39}
\newcommand{\sigTwoLo}{0.14}
\newcommand{\sigTwoHi}{0.15}
\newcommand{\sdssWoff}{0.26}
\newcommand{\WsdssMedL}{3150}
\newcommand{\WlrdMedL}{1710}

\newcommand{\QSOsn}{3.4}
\newcommand{\QSOW}{2107}
\newcommand{\QSOmu}{7.5}
\newcommand{\QSOmuErr}{1.5}
\newcommand{\QSOresCor}{0.17}
\newcommand{\QSOresErr}{0.08}
\newcommand{\QSOcompNar}{560}
\newcommand{\QSOcompBrd}{2000}
\newcommand{\QSOWfix}{1223}
\newcommand{\QSOsnFix}{20.9}
\newcommand{\QSOresFix}{\ensuremath{-}0.08}
\newcommand{\QSOresFixErr}{0.03}
\newcommand{\FsigNarOIIIqso}{0.072}
\newcommand{\NLRDqso}{30}

\newcommand{\vKnorm}{1493}
\newcommand{\kappaES}{0.398}
\newcommand{\NPA}{18}
\newcommand{\bPA}{0.21}
\newcommand{\bPAlo}{0.10}
\newcommand{\bPAhi}{0.31}
\newcommand{\cPA}{0.34}
\newcommand{\cPAlo}{\ensuremath{-}0.55}
\newcommand{\cPAhi}{1.28}
\newcommand{\hpdPA}{83}
\newcommand{\sqPA}{0.073}
\newcommand{\NPB}{15}
\newcommand{\bPB}{0.15}
\newcommand{\bPBlo}{0.02}
\newcommand{\bPBhi}{0.26}
\newcommand{\cPB}{0.57}
\newcommand{\cPBlo}{\ensuremath{-}0.59}
\newcommand{\cPBhi}{1.73}
\newcommand{\sqPB}{0.081}
\newcommand{\NPC}{nine}
\newcommand{\bPC}{0.40}
\newcommand{\bPClo}{0.21}
\newcommand{\bPChi}{0.57}
\newcommand{\cPC}{\ensuremath{-}0.10}
\newcommand{\cPClo}{\ensuremath{-}0.69}
\newcommand{\cPChi}{0.53}
\newcommand{\hpdPC}{98}
\newcommand{\sqPC}{0.115}
\newcommand{\NPdrop}{five}
\newcommand{\NPAall}{23}
\newcommand{\NPBall}{18}
\newcommand{\sxDropMin}{0.4}
\newcommand{\sxDropMax}{2.0}
\newcommand{\sxKeepMax}{0.18}
\newcommand{\hierDb}{0.01}
\newcommand{\NPE}{five}
\newcommand{\sqPElo}{0.019}
\newcommand{\sqPEhi}{0.346}
\newcommand{\NPCbx}{four}
\newcommand{\NdGall}{146}
\newcommand{\NdGbb}{47}
\newcommand{\NdGfit}{14}
\newcommand{\ratLo}{2.5}
\newcommand{\ratHi}{3.0}
\newcommand{\sqLo}{0.07}
\newcommand{\sqHi}{0.08}
\newcommand{\TminP}{3600}
\newcommand{\TmaxP}{6400}
\newcommand{\LphMin}{43.3}
\newcommand{\LphMax}{44.7}
\newcommand{\sdLphHaLo}{0.22}
\newcommand{\sdLphHaHi}{0.23}
\newcommand{\lineB}{0.2}
\newcommand{\lineCLo}{0.2}
\newcommand{\lineCHi}{0.7}
\newcommand{\NJE}{five}
\newcommand{\bpJE}{\ensuremath{-}0.01}
\newcommand{\bpJElo}{\ensuremath{-}0.64}
\newcommand{\bpJEhi}{0.60}
\newcommand{\simMedJE}{0.4}
\newcommand{\NJA}{13}
\newcommand{\bpJA}{\ensuremath{-}0.01}
\newcommand{\bpJAlo}{\ensuremath{-}0.16}
\newcommand{\bpJAhi}{0.15}
\newcommand{\simPctJA}{4}
\newcommand{\simMedJA}{0.14}
\newcommand{\slJA}{0.031}
\newcommand{\soJA}{0.040}
\newcommand{\NJB}{15}
\newcommand{\bpJB}{\ensuremath{-}0.05}
\newcommand{\bpJBlo}{\ensuremath{-}0.25}
\newcommand{\bpJBhi}{0.14}
\newcommand{\simPctJB}{1}
\newcommand{\simMedJB}{0.21}
\newcommand{\slJB}{0.057}
\newcommand{\soJB}{0.081}
\newcommand{\powTen}{20}
\newcommand{\powLo}{70}
\newcommand{\powHi}{80}
\newcommand{\kRing}{2}
\newcommand{\kIso}{1.36}
\newcommand{\kVir}{0.96}
\newcommand{\lamVirLo}{0.10}
\newcommand{\lamVirHi}{0.15}
\newcommand{\lamIsoLo}{0.21}
\newcommand{\lamIsoHi}{0.30}
\newcommand{\lamRingLo}{0.45}
\newcommand{\lamRingHi}{0.64}
\newcommand{\mEddLo}{1.6}
\newcommand{\mEddHi}{10}
\newcommand{\mEddStdLo}{3.4}
\newcommand{\brdVirLo}{2.6}
\newcommand{\brdVirHi}{3.1}
\newcommand{\brdIsoLo}{1.8}
\newcommand{\brdIsoHi}{2.2}
\newcommand{\brdRingLo}{1.25}
\newcommand{\brdRingHi}{1.5}
\newcommand{\deltaKthreeLo}{0.003}
\newcommand{\deltaKthreeHi}{0.030}
\newcommand{\deltaKfourMax}{0.30}

\newcommand{\lamEscKfourMax}{1.42}

\newcommand{\sigIncA}{0.053}
\newcommand{\NIncB}{35}
\newcommand{\sigIncB}{0.059}
\newcommand{\NIncD}{42}
\newcommand{\sigIncD}{0.069}
\newcommand{\chiHiB}{66}

\newcommand{\ratNpB}{2.55}
\newcommand{\ratNpE}{2.57}
\newcommand{\sqMeasLo}{0.07}
\newcommand{\sqMeasHi}{0.10}
\newcommand{\ratExLo}{1.4}
\newcommand{\ratWingHi}{2.0}

\newcommand{\NSNtwenty}{nine}
\newcommand{\sigSNtwenty}{0.065}
\newcommand{\floorDex}{0.15}
\newcommand{\kappaSolar}{0.34}
\newcommand{\vKdrop}{8}

\newcommand{\NSsgDraw}{300}
\newcommand{\NSsg}{224}
\newcommand{\Ssg}{0.123}
\newcommand{\Ssglo}{0.112}
\newcommand{\Ssghi}{0.136}
\newcommand{\SsgTwo}{0.12}
\newcommand{\SsgCat}{0.185}
\newcommand{\sdssRatSame}{2.3}
\newcommand{\fKepNpLo}{38}
\newcommand{\fKepNpHi}{44}
\newcommand{\fEscNpLo}{23}
\newcommand{\fEscNpHi}{28}
\newcommand{\hierSig}{0.054}
\newcommand{\hierB}{0.17}

\newcommand{\twoFbLo}{0.18}
\newcommand{\twoFbHi}{0.34}
\newcommand{\sigLamMax}{0.15}
\newcommand{\oriThin}{0.21}
\newcommand{\oriHalf}{0.081}
\newcommand{\oriSeven}{0.056}
\newcommand{\oriOne}{0.035}
\newcommand{\oriThinRatio}{2.8}
\newcommand{\rhoThin}{0.93}
\newcommand{\sigLamIso}{0.13}
\newcommand{\oriEdgeWide}{0.042}
\newcommand{\oriEdge}{0.018}
\newcommand{\sigLamEdgeWide}{0.12}
\newcommand{\sigLamEdge}{0.14}
\newcommand{\sigDustMax}{1.1}
\newcommand{\sigDustNoTop}{1.8}
\newcommand{\RLlo}{0.13}
\newcommand{\RLhi}{0.19}
\newcommand{\RLtermLo}{0.065}
\newcommand{\RLtermHi}{0.095}
\newcommand{\virW}{2.06}
\newcommand{\virL}{0.47}
\newcommand{\sigMvir}{0.16}
\newcommand{\sigMvirMed}{0.11}
\newcommand{\sigMvirRep}{0.12}
\newcommand{\sigMvirCons}{0.21}

\newcommand{\rusA}{428}
\newcommand{\rusB}{370}
\newcommand{\DHalf}{0.37}
\newcommand{\DOne}{0.54}
\newcommand{\DTwo}{0.70}
\newcommand{\DThree}{0.78}
\newcommand{\sigTauHalf}{0.20}
\newcommand{\sigTauOne}{0.14}
\newcommand{\sigTauTwo}{0.11}
\newcommand{\sigTauThree}{0.10}
\newcommand{\sigTeMax}{0.15}
\newcommand{\sigRHalf}{0.41}
\newcommand{\sigROne}{0.28}
\newcommand{\sigRTwo}{0.21}
\newcommand{\sigRThree}{0.19}
\newcommand{\tauPhLo}{0.09}
\newcommand{\coreLo}{500}
\newcommand{\coreHi}{1000}
\newcommand{\DconvHalfLo}{0.21}
\newcommand{\DconvHalfHi}{0.29}
\newcommand{\sigTauConvHalfLo}{0.26}
\newcommand{\sigTauConvHalfHi}{0.35}
\newcommand{\DconvHiLo}{0.56}
\newcommand{\DconvHiHi}{0.75}
\newcommand{\sigTauConvHiLo}{0.10}
\newcommand{\sigTauConvHiHi}{0.13}
\newcommand{\sigRConvHiLo}{0.20}
\newcommand{\sigRConvHiHi}{0.26}
\newcommand{\rtHalfBoostMax}{20}
\newcommand{\DrtHiLo}{0.94}
\newcommand{\DrtHiHi}{1.61}

\newcommand{\Hnot}{67.7}
\newcommand{\OmM}{0.310}
\newcommand{\Hcat}{70}
\newcommand{\OmCat}{0.3}
\newcommand{\dCosmoL}{0.02}
\newcommand{\dCosmoVK}{0.005}
\newcommand{\fitWin}{5000}
\newcommand{\errFloor}{2}
\newcommand{\wnarLo}{30}
\newcommand{\wnarHi}{600}
\newcommand{\wbroadLo}{350}
\newcommand{\wbroadHi}{12{,}000}
\newcommand{\tauMax}{10}
\newcommand{\wabsLo}{50}
\newcommand{\wabsHi}{1500}
\newcommand{\nBootNp}{200}
\newcommand{\winWide}{6800}
\newcommand{\NBICgauss}{four}
\newcommand{\NBIClor}{16}
\newcommand{\NBICexp}{nine}
\newcommand{\dWge}{0.264}
\newcommand{\DsigBic}{0.111}
\newcommand{\rResSN}{0.08}
\newcommand{\pResSN}{0.69}
\newcommand{\FsigWin}{0.062}
\newcommand{\FsigLowSN}{0.059}
\newcommand{\Ssgb}{0.10}
\newcommand{\fEscInfLo}{24}
\newcommand{\fEscInfHi}{32}
\newcommand{\recLG}{0.071}
\newcommand{\hitTenAll}{three}
\newcommand{\nullDenThree}{300}
\newcommand{\bNullHi}{0.07}
\newcommand{\hitSlopeNull}{none}
\newcommand{\kIsoExp}{0.57}
\newcommand{\lamIsoExpLo}{0.16}
\newcommand{\lamIsoExpHi}{0.17}
\newcommand{\ratTightLo}{1.3}
\newcommand{\ratTightHi}{1.4}

%% file: tables/Table1_stub.tex
\begin{deluxetable*}{llcccllcccc}
\tabletypesize{\footnotesize}
\tablewidth{0pt}
\tablecaption{The sample\label{tab:sample}}
\tablehead{\colhead{Source} & \colhead{Class} & \colhead{R.A.} & \colhead{Decl.} & \colhead{$z$} & \colhead{Program} & \colhead{Grating} & \colhead{Label} & \colhead{FWHM} & \colhead{$\log L_{\rm H\alpha}$} & \colhead{S/N} \\
\colhead{} & \colhead{} & \colhead{(deg)} & \colhead{(deg)} & \colhead{} & \colhead{} & \colhead{} & \colhead{} & \colhead{(km s$^{-1}$)} & \colhead{(erg s$^{-1}$)} & \colhead{}}
\startdata
RUBIES-EGS-42046 & LRD & 214.79537 & $52.78885$ & 5.276 & 8410 & G395H & T25 & $3185\pm48$ & $43.55\pm0.01$ & 71.3 \\
CEERS-746 & LRD & 214.80915 & $52.86848$ & 5.623 & 4106 & G395M & T25 & $1777\pm45$ & $42.27\pm0.01$ & 38.1 \\
RUBIES-UDS-119957 & LRD & 34.26891 & $-5.17672$ & 4.149 & 4233 & G395M & T25 & $1811\pm113$ & $42.20\pm0.03$ & 14.1 \\
Skyfire-299 & LRD & 214.80575 & $52.87805$ & 5.623 & 4106 & G395M & K26 & $1989\pm72$ & $42.29\pm0.02$ & 19.5 \\
RUBIES-EGS-15825 & other & 215.07926 & $52.93425$ & 3.666 & 4233 & G395M & T25 & $3612\pm184$ & $42.75\pm0.03$ & 14.9 \\
\enddata
\tablecomments{FWHM is the intrinsic width of the single broad Gaussian and $L_{\rm H\alpha}$ the luminosity of that component before absorption, without dust correction. Uncertainties are $1\sigma$, conditional on the adopted model. Labels: T25 = \citet{Taylor2025}; K26 = \citet{Kocevski2026}; H25 = \citet{Hviding2025}. Five rows are shown. The full table of 44 sources, with the five other width measures and all covariances, is published in machine-readable form.}
\end{deluxetable*}

%% file: tables/Table2_fits.tex
\centerwidetable
\begin{deluxetable*}{lclclcl}
\tabletypesize{\scriptsize}
\tabcolsep=1.7pt
\tablewidth{0pt}
\tablecaption{Width--luminosity fits\label{tab:fits}}
\tablehead{\colhead{} & \multicolumn{3}{c}{LRDs} & \multicolumn{2}{c}{Other broad-line sources} & \colhead{} \\ \cline{2-4} \cline{5-6}
\colhead{Width measure or variant} & \colhead{$N$} & \colhead{$b$} & \colhead{$\sigma_{\rm int}$ (dex)} & \colhead{$N$} & \colhead{$\sigma_{\rm int}$ (dex)} & \colhead{$\Delta\sigma_{\rm int}$ (dex)}}
\startdata
\cutinhead{This work: six measures of width}
Single-Gaussian FWHM\tablenotemark{a} & 29 & $0.17$ ($0.12$--$0.22$) & $0.054$ ($0.038$--$0.080$) & 15 & $0.154$ ($0.104$--$0.252$) & $-0.099$ ($-0.199$ to $-0.043$) \\
Interquartile width $W_{50}$ & 29 & $0.13$ ($0.08$--$0.18$) & $0.051$ ($0.037$--$0.073$) & 15 & $0.147$ ($0.101$--$0.238$) & $-0.095$ ($-0.187$ to $-0.045$) \\
Exponential FWHM & 29 & $0.14$ ($0.08$--$0.19$) & $0.051$ ($0.029$--$0.085$) & 15 & $0.169$ ($0.102$--$0.296$) & $-0.117$ ($-0.245$ to $-0.042$) \\
Lorentzian FWHM & 29 & $0.16$ ($0.10$--$0.23$) & $0.062$ ($0.039$--$0.099$) & 15 & $0.271$ ($0.173$--$0.456$) & $-0.207$ ($-0.394$ to $-0.104$) \\
Nonparametric FWHM & 29 & $0.15$ ($0.08$--$0.22$) & $0.076$ ($0.056$--$0.109$) & 15 & $0.155$ ($0.106$--$0.250$) & $-0.078$ ($-0.175$ to $-0.020$) \\
Two-Gaussian FWHM & 29 & $0.09$ ($-0.02$ to $0.21$) & $0.118$ ($0.082$--$0.176$) & 15 & $0.214$ ($0.123$--$0.402$) & $-0.095$ ($-0.285$ to $0.013$) \\
\cutinhead{This work: single-Gaussian FWHM, error-floor and sample variants}
Repeat floor, same grating ($+0.044$ dex)\tablenotemark{a} & 29 & $0.17$ ($0.12$--$0.22$) & $0.027$ ($0.002$--$0.063$) & 15 & $0.147$ ($0.093$--$0.248$) & $-0.120$ ($-0.224$ to $-0.054$) \\
Repeat floor, all pairs ($+0.060$ dex)\tablenotemark{a} & 29 & $0.17$ ($0.11$--$0.23$) & $0.018$ ($0.001$--$0.056$) & 15 & $0.140$ ($0.083$--$0.243$) & $-0.121$ ($-0.226$ to $-0.053$) \\
LSF width $\times0.5$\tablenotemark{a} & 29 & $0.19$ ($0.13$--$0.26$) & $0.069$ ($0.049$--$0.100$) & 15 & $0.157$ ($0.106$--$0.258$) & $-0.088$ ($-0.190$ to $-0.028$) \\
Three brightest LRDs removed & 26 & $0.09$ ($0.01$--$0.19$) & $0.040$ ($0.020$--$0.069$) & 15 & $0.154$ ($0.104$--$0.252$) & $-0.114$ ($-0.214$ to $-0.056$) \\
LRDs without Balmer absorption & \NNoAbs{} & $\FbNoAbs{}$ ($\FbNoAbslo{}$--$\FbNoAbshi{}$) & $\FsigNoAbs{}$ ($\FsigNoAbslo{}$--$\FsigNoAbshi{}$) & \NOther{} & $\Fsigo{}$ ($\Fsigolo{}$--$\Fsigohi{}$) & $\FdsigNoAbs{}$ ($\FdsigNoAbslo{}$ to $\FdsigNoAbshi{}$) \\
Narrow-line FWHM $<\wnarHi{}$~\kms{}\,\tablenotemark{a} & \NLRDNarCut{} & $\FbNarCut{}$ ($\FbNarCutlo{}$--$\FbNarCuthi{}$) & $\FsigNarCut{}$ ($\FsigNarCutlo{}$--$\FsigNarCuthi{}$) & \NOther{} & $\Fsigo{}$ ($\Fsigolo{}$--$\Fsigohi{}$) & $\FdsigNarCut{}$ ($\FdsigNarCutlo{}$ to $\FdsigNarCuthi{}$) \\
Narrow-line FWHM fixed to \wnarDef{}~\kms{}\,\tablenotemark{a} & \NLRD{} & $\FbNarFix{}$ ($\FbNarFixlo{}$--$\FbNarFixhi{}$) & $\FsigNarFix{}$ ($\FsigNarFixlo{}$--$\FsigNarFixhi{}$) & \NOther{} & $\FsigoNarFix{}$ ($\FsigoNarFixlo{}$--$\FsigoNarFixhi{}$) & $\FdsigNarFix{}$ ($\FdsigNarFixlo{}$ to $\FdsigNarFixhi{}$) \\
Narrow-line FWHM fixed to [\ion{O}{3}]\tablenotemark{a} & \NLRD{} & $\FbNarOIII{}$ ($\FbNarOIIIlo{}$--$\FbNarOIIIhi{}$) & $\FsigNarOIII{}$ ($\FsigNarOIIIlo{}$--$\FsigNarOIIIhi{}$) & \NOther{} & $\FsigoNarOIII{}$ ($\FsigoNarOIIIlo{}$--$\FsigoNarOIIIhi{}$) & $\FdsigNarOIII{}$ ($\FdsigNarOIIIlo{}$ to $\FdsigNarOIIIhi{}$) \\
{}[\ion{O}{3}]-anchored, absorption rule re-applied\tablenotemark{a,e} & 27 & \nodata & $0.075$ ($0.055$--$0.105$) & \nodata & \nodata & \nodata \\
Fit window $\pm\winWider{}$~\kms{}, [\ion{S}{2}] masked & 29 & $0.17$ ($0.09$--$0.24$) & $0.083$ ($0.059$--$0.119$) & 15 & $0.182$ ($0.124$--$0.295$) & $-0.098$ ($-0.213$ to $-0.031$) \\
Lowest-S/N passing spectrum\tablenotemark{a} & 29 & \nodata & $0.059$ ($0.039$--$0.088$) & \nodata & \nodata & \nodata \\
In-band pre-cut ${\rm S/N}\geq\cutSNthree{}$ included\tablenotemark{a} & \NLRDSNthree{} & $\FbSNthree{}$ ($\FbSNthreelo{}$--$\FbSNthreehi{}$) & $\FsigSNthree{}$ ($\FsigSNthreelo{}$--$\FsigSNthreehi{}$) & \nodata & \nodata & \nodata \\
Two widest other sources removed & 29 & $0.17$ ($0.12$--$0.22$) & $0.054$ ($0.038$--$0.080$) & 13 & $0.097$ ($0.061$--$0.172$) & $-0.043$ ($-0.119$ to $0.002$) \\
\cutinhead{Table of \citet{BaccusXu2026}}
All records\tablenotemark{b} & 48 & $0.25$ ($0.20$--$0.31$) & $0.058$ ($0.045$--$0.077$) & 97 & $0.144$ ($0.122$--$0.172$) & $-0.085$ ($-0.117$ to $-0.057$) \\
Shared records removed & 28 & $0.25$ ($0.17$--$0.34$) & $0.075$ ($0.054$--$0.109$) & 88 & $0.139$ ($0.117$--$0.168$) & $-0.064$ ($-0.100$ to $-0.024$) \\
Matched pairs\tablenotemark{c} & 36 & \nodata & \nodata & 36 & \nodata & $-0.065$ ($-0.105$ to $-0.024$) \\
\cutinhead{SDSS DR7 quasars \citep{Shen2011}}
Multi-Gaussian FWHM (catalog) & \nodata & \nodata & \nodata & 3755 & $0.208$ ($0.203$--$0.213$) & \nodata \\
Single-Gaussian FWHM, refitted\tablenotemark{d} & \nodata & \nodata & \nodata & 224 & $0.123$ ($0.112$--$0.136$) & \nodata \\
\enddata
\tablecomments{Posterior medians, with equal-tailed 95\% intervals in parentheses. $b$ is the slope of $\log{\rm FWHM}$ on $\log L$, and $\Delta\sigma_{\rm int}$ is the scatter of the LRDs minus that of the other sources. Each width measure is regressed on the luminosity and covariance of its own fit ($L_{\rm H\alpha}$ for our sample and SDSS; $L_{\rm bol}$ for \citealt{BaccusXu2026}).}
\tablenotetext{a}{One-sided 95\% upper limits (95th posterior percentiles) on LRD $\sigma_{\rm int}$ are \FsigUp{}~dex under formal errors, \FsigFlrSameUp{}~dex with the $+\flrDupSameG{}$~dex same-grating repeat-spectrum floor added in quadrature to $C_{yy}$, \FsigFlrAllUp{}~dex with the $+\flrDupAll{}$~dex all-pairs floor, \FsigLsfUp{}~dex for LSF width $\times0.5$, \FsigNarCutUp{}~dex for narrow-line FWHM $<\wnarHi{}$~\kms{}, \FsigNarFixUp{}~dex for narrow-line FWHM fixed to \wnarDef{}~\kms{}, \FsigNarOIIIUp{}~dex when anchored to [\ion{O}{3}]~$\lambda5007$ (\wnarDef{}~\kms{} where unmeasured; [\ion{O}{3}] hits the \wnarLo{}~\kms{} lower bound in \NOIIIUnres{} of the \NOIIIMeas{} LRDs where it is measured, and fixing all \NLRD{} to \wnarLo{}~\kms{} gives $\sigma_{\rm int}=\FsigNarMin{}$~dex, $<\FsigNarMinUp{}$~dex), \FsigOIIIqcUp{}~dex when the absorption-bound rule is re-applied to the [\ion{O}{3}]-anchored fits, \FsigLowSNUp{}~dex for the lowest-S/N passing spectra, and \FsigSNthreeUp{}~dex with in-band failed ${\rm S/N}\geq\cutSNthree{}$ pre-cut LRD spectra included (Sections~\ref{sec:relation} and \ref{sec:tests}). Random-effects fits to all repeat spectra give \reAllUp{}~dex (additive floor) and \reMultUp{}~dex (multiplicative).}
\tablenotetext{b}{The \NXLRD{} LRD records correspond to \NXLRDuniq{} unique objects on the sky; deduplicating to one record per object gives $\sigma_{\rm int}=\XsigDedup{}$~dex (\XsigDeduplo{}--\XsigDeduphi{}).}
\tablenotetext{c}{Pairs matched in grating, redshift ($|\Delta z|<0.5$), and luminosity ($|\Delta\log L|<0.2$). The entry is the difference of the residual standard deviations, with its bootstrap interval.}
\tablenotetext{d}{A random 300 of the 3755 quasars refitted with the procedure of Section~\ref{sec:fit}; 224 pass the rules of Section~\ref{sec:sample}, giving slope $b=0.10$ ($0.05$--$0.15$).}
\tablenotetext{e}{Two LRDs whose absorption optical depth reaches its bound in the [\ion{O}{3}]-anchored refit are removed; the shift from the 29-object row is evaluated with the baseline measurement covariances.}
\end{deluxetable*}

%% file: tables/Table3_readings.tex
\begin{deluxetable*}{lll}
\tabletypesize{\scriptsize}
\tablewidth{0pt}
\tablecaption{Empirical constraints on the two readings of LRD broad \Ha{} widths\label{tab:readings}}
\tablehead{\colhead{Empirical property} & \colhead{Bound orbital motion (${\rm FWHM}=f_{\rm vir}^{-1/2}\sqrt{GM/R}$)} & \colhead{Scattering envelope ($\lambda\geq1$, $R\geq\Rph$)}}
\startdata
\parbox[t]{0.26\textwidth}{\raggedright Velocity scale \& wings: ${\rm FWHM}/\vK=\ratLo{}\text{--}\ratHi{}$; \fEscNpLo{}--\fEscNpHi{}\% flux at $|v|>\sqrt{2}\,\vK$\strut} & \parbox[t]{0.335\textwidth}{\raggedright $\lambda\leq\lamVirLo{}\text{--}\lamIsoHi{}$ ($f_{\rm vir}=1.08\text{--}0.54$) or $\leq\lamRingHi{}$ ($f_{\rm vir}=0.25$, edge-on ring) for $R\geq\Rph$\strut} & \parbox[t]{0.335\textwidth}{\raggedright Core broadened $\brdVirLo{}\text{--}\brdVirHi{}\times$ ($f_{\rm vir}=1.08$) to $\brdRingLo{}\text{--}\brdRingHi{}\times$ ($f_{\rm vir}=0.25$) at $\lambda=1$ ($\tau_e\simeq2$--3, $T_e\sim10^{4}$~K)\strut} \\[3pt]
\parbox[t]{0.26\textwidth}{\raggedright Width scatter at fixed $\LHa$: $\sigint<\FsigUp{}$~dex ($<\FsigCons{}$~dex conservative)\strut} & \parbox[t]{0.335\textwidth}{\raggedright $\sigma_{M/R}\leq\sigLamMax{}$~dex; $H/R\simeq1$ ($\sigma_{M/R}\leq\sigLamIso{}$~dex) or $i>45^\circ$ ($\leq\sigLamEdgeWide{}\text{--}\sigLamEdge{}$~dex); $\sigma_A\leq\sigDustMax{}$~mag\strut} & \parbox[t]{0.335\textwidth}{\raggedright $\sigma_{\log\tau_e}\leq\sigTauThree{}\text{--}\sigTauTwo{}$~dex (\sigTauConvHiLo{}--\sigTauConvHiHi{}~dex with core) or $\sigma_{\log T_e}\leq\sigTeMax{}$~dex; $\sigma_{\log R}\lesssim\sigRThree{}\text{--}\sigRTwo{}$~dex\strut} \\[3pt]
\parbox[t]{0.26\textwidth}{\raggedright Width--luminosity slope: $b=\Fb{}$ (\FbNarFix{}--\FbNarOIII{} narrow-corrected)\strut} & \parbox[t]{0.335\textwidth}{\raggedright $M/R\propto\LHa^{2b}=\LHa^{\twoFbLo{}\text{--}\twoFbHi{}}$\strut} & \parbox[t]{0.335\textwidth}{\raggedright ${\rm FWHM}\propto\LHa^{\shellSlopeLo{}\text{--}\shellSlopeHi{}}$ at $\tau_e\simeq2$--3 (narrow-corrected $b$ at or below lower edge)\strut} \\[3pt]
\parbox[t]{0.26\textwidth}{\raggedright Null $\Lph$ trend at fixed $\LHa$: partial slope $\bpJA{}$ ($\sigma[\log(\Lph/\LHa)]=\sdLphHaLo{}\text{--}\sdLphHaHi{}$~dex)\strut} & \parbox[t]{0.335\textwidth}{\raggedright Photoionized BLR ($R$ tracks $\LHa$, not $\Rph$)\strut} & \parbox[t]{0.335\textwidth}{\raggedright Cospatial recombination and scattering ($\LHa\propto R\,\tau_e^{2}$)\strut} \\[3pt]
\parbox[t]{0.26\textwidth}{\raggedright Single-epoch virial mass: $\Mvir\propto\LHa^{\virL{}}\,{\rm FWHM}^{\virW{}}$\strut} & \parbox[t]{0.335\textwidth}{\raggedright $\sigma_{\Mvir}<\sigMvir{}$~dex ($<\sigMvirCons{}$~dex conservative); $M({<}R)=(\mEddStdLo{}\text{--}\mEddHi{})\,(R/\Rph)\,M_{\rm Edd}(\Lph)$ ($\mEddLo{}\times$ for $f_{\rm vir}=0.25$)\strut} & \parbox[t]{0.335\textwidth}{\raggedright $\sigma_{\Mvir}<\sigMvir{}$~dex ($<\sigMvirCons{}$~dex conservative); $M_{\rm BH}=\lambda^{-1}M_{\rm Edd}(\Lph)$\strut} \\
\enddata
\tablecomments{Ceilings in columns 2--3 use the baseline 95\% limit $\sigint=\FsigUp{}$~dex (loosened by \narCeilLoosenPct{}\% at the conservative \FsigCons{}~dex ceiling, tightened by 25\% with the same-grating repeat floor), with empirical reference radius $\Rph\equiv(\Lph/4\pi\sigma_{\rm SB}\Teff^{4})^{1/2}$ and $M_{\rm Edd}(\Lph)=\kappa_{\rm es}\Lph/(4\pi G c)=8\times10^{5}\,\Msun\,(\Lph/10^{44}\,\ergs)$. In column~2, rows~2--3 hold for any bound virial BLR under the Section~\ref{sec:budget} independence assumptions, row~4 lists natural physical couplings for the flat $\Lph$ residuals, and $M({<}R)/M_{\rm Edd}(\Lph)\propto R/\Rph$ (row~5) holds generally; only the $R\geq\Rph$ occultation bound enforcing $\lambda\leq\lamVirLo{}\text{--}\lamIsoHi{}$ (row~1) assumes a spherical photosphere; flattened disks avoid spherical-envelope tension either at $\lambda\ll1$ with $R\simeq(\lamVirLo{}\text{--}\lamIsoHi{})\,\lambda^{-1}\Rph\gtrsim\Rph$ or by exposing $R<\Rph$ to lift the $\lambda$ ceiling (Section~\ref{sec:alternatives}). Column~3 rows~2--4 assume the isothermal-shell wing calibration and cospatial $\LHa=K R\,\tau_e^{2}$ without parameter covariance (Section~\ref{sec:readings}).}
\end{deluxetable*}

%% file: tables/TableB1_injections.tex
\begin{deluxetable*}{llccccc}
\tabletypesize{\scriptsize}
\tabcolsep=2.2pt
\tablewidth{0pt}
\tablecaption{Injection--recovery tests on the LRD noise and LSF templates\label{tab:injections}}
\tablehead{\colhead{Injected profile \& relation} & \colhead{Width measure} & \colhead{$\sigma_{\rm parent}$} & \colhead{Injected $b$} & \colhead{Recovered $\sigma_{\rm int}$ (dex)} & \colhead{Trials $\leq\Fsig{}$~dex} & \colhead{Recovered $b$}}
\startdata
\cutinhead{Cross-profile matrix at fiducial scatter}
Gaussian, sloped & Single-Gaussian & $0.05$ & $0.17$ & $0.052$ ($0.040$--$0.061$) & 62/100 & \nodata \\
Gaussian, sloped & Interquartile $W_{50}$ & $0.05$ & $0.17$ & $0.056$ ($0.048$--$0.068$) & 29/100 & \nodata \\
Gaussian, sloped & Exponential & $0.05$ & $0.17$ & $0.102$ ($0.087$--$0.128$) & 1/100 & \nodata \\
Gaussian, sloped & Lorentzian & $0.05$ & $0.17$ & $0.135$ ($0.099$--$0.177$) & 3/100 & \nodata \\
Exponential, sloped & Single-Gaussian & $0.05$ & $0.17$ & $0.055$ ($0.043$--$0.070$) & 44/100 & \nodata \\
Exponential, sloped & Interquartile $W_{50}$ & $0.05$ & $0.17$ & $0.057$ ($0.048$--$0.069$) & 22/100 & \nodata \\
Exponential, sloped & Exponential & $0.05$ & $0.17$ & $0.039$ ($0.024$--$0.058$) & 68/100 & \nodata \\
Exponential, sloped & Lorentzian & $0.05$ & $0.17$ & $0.047$ ($0.030$--$0.068$) & 70/100 & \nodata \\
Lorentzian, matched locus & Single-Gaussian & $0.05$ & $0.15$ & $0.071$ ($0.055$--$0.097$) & 14/100 & $0.11$ ($0.08$--$0.15$) \\
\cutinhead{Parent-scatter and truncation tests}
Gaussian, observed locus & Single-Gaussian & $0.08$ & $0.17$ & $0.083$ ($0.067$--$0.095$) & 3/100 & $0.17$ ($0.12$--$0.20$) \\
Exponential, matched locus & Single-Gaussian & $0.08$ & $0.18$ & $0.077$ ($0.063$--$0.100$) & 4/100 & $0.13$ ($0.09$--$0.17$) \\
Lorentzian, matched locus & Single-Gaussian & $0.08$ & $0.15$ & $0.088$ ($0.069$--$0.109$) & 2/100 & $0.11$ ($0.06$--$0.14$) \\
Gaussian, observed locus & Single-Gaussian & $0.10$ & $0.17$ & $0.102$ ($0.085$--$0.119$) & 0/100 & $0.18$ ($0.13$--$0.22$) \\
Exponential, matched locus & Single-Gaussian & $0.10$ & $0.18$ & $0.094$ ($0.072$--$0.118$) & 1/100 & $0.12$ ($0.09$--$0.17$) \\
Lorentzian, matched locus & Single-Gaussian & $0.10$ & $0.15$ & $0.093$ ($0.075$--$0.115$) & 2/100 & $0.12$ ($0.07$--$0.16$) \\
Exponential, $+0.32$~dex wider & Single-Gaussian & $0.10$ & $0.18$ & $0.074$ ($0.060$--$0.095$) & 6/100 & $0.14$ ($0.10$--$0.18$) \\
Gaussian, lum.-independent & Single-Gaussian & $0.15$ & $0.00$ & $0.145$ ($0.125$--$0.170$) & 0/100 & \nodata \\
Exponential, lum.-independent & Single-Gaussian & $0.15$ & $0.00$ & $0.098$ ($0.080$--$0.121$) & 0/100 & \nodata \\
\cutinhead{Null-slope tests ($b=0$)}
Exponential, matched width & Single-Gaussian & $0.05$ & $0.00$ & $0.059$ ($0.045$--$0.073$) & 41/100 & $0.02$ ($-0.01$ to $0.04$) \\
Lorentzian, matched width & Single-Gaussian & $0.05$ & $0.00$ & $0.068$ ($0.054$--$0.091$) & 18/100 & $0.02$ ($-0.01$ to $0.04$) \\
\cutinhead{Full in-band parent templates (\NParentInBand{} LRDs)\tablenotemark{a}}
Gaussian, observed locus & Single-Gaussian & $0.05$ & $0.17$ & $\recParFiveG{}$ ($\recParFiveGlo{}$--$\recParFiveGhi{}$) & \hitParFiveGnum{} & $\bParFiveG{}$ ($\bParFiveGlo{}$--$\bParFiveGhi{}$) \\
Exponential, matched locus & Single-Gaussian & $0.05$ & $0.18$ & $\recParFiveE{}$ ($\recParFiveElo{}$--$\recParFiveEhi{}$) & \hitParFiveEnum{} & $\bParFiveE{}$ ($\bParFiveElo{}$--$\bParFiveEhi{}$) \\
Gaussian, observed locus & Single-Gaussian & $0.10$ & $0.17$ & $\recParTenG{}$ ($\recParTenGlo{}$--$\recParTenGhi{}$) & \hitParTenGnum{} & $\bParTenG{}$ ($\bParTenGlo{}$--$\bParTenGhi{}$) \\
Exponential, matched locus & Single-Gaussian & $0.10$ & $0.18$ & $\recParTenE{}$ ($\recParTenElo{}$--$\recParTenEhi{}$) & \hitParTenEnum{} & $\bParTenE{}$ ($\bParTenElo{}$--$\bParTenEhi{}$) \\
\enddata
\tablecomments{Each row summarizes \NReal{} simulated samples generated on the pixel grids, noise vectors, and grating LSFs of the observed LRD spectra (\NLRD{} surviving templates in the first three blocks; all \NParentInBand{} in-band parent templates in the bottom block), refitted with the pipeline of Section~\ref{sec:fit}, and filtered by the quality rules of Section~\ref{sec:sample}. Recovered $\sigma_{\rm int}$ and $b$ list the median (16th--84th percentiles) of the Bayesian posterior medians across the \NReal{} realizations. In the null-slope ($b=0$) rows, \hitSlopeNull{} of \nullDen{} realizations yield a posterior-median slope as steep as $b=\Fb{}$ (97.5th percentile $\leq\bNullHi{}$).}
\tablenotetext{a}{Across the \NParentInBand{}-template suites, a median of \NParSelG{} (\NParSelGlo{}--\NParSelGhi{}, Gaussian) and \NParSelE{} (\NParSelElo{}--\NParSelEhi{}, exponential) sources pass Section~\ref{sec:sample} per realization; replacing refitted/published $\LHa$ with pipeline-fitted $\LHa$ on all failed templates yields $\sigma_{\rm int}\leq\Fsig{}$~dex in \hitParOwnTenGnum{} and \hitParOwnTenEnum{} realizations at $\sigma_{\rm parent}=\simA{}$~dex.}
\end{deluxetable*}